\documentclass[fleqn,usenatbib]{mnras}

\usepackage{newtxtext,newtxmath}
\usepackage{float}
\usepackage[T1]{fontenc}

\DeclareRobustCommand{\VAN}[3]{#2}
\let\VANthebibliography\thebibliography
\def\thebibliography{\DeclareRobustCommand{\VAN}[3]{##3}\VANthebibliography}

\usepackage{graphicx}	% Including figure files
\usepackage{amsmath}	% Advanced maths commands
\usepackage{xcolor}

\usepackage{hyperref}

\hypersetup{
colorlinks=true,
linkcolor=blue,
filecolor=magenta,
urlcolor=cyan,
}

\title[WDM subhalo distribution]{Extending the unified subhalo model to warm dark matter}
\author[He et al.]{
Feihong He,$^{1,2,3}$
Jiaxin Han,$^{1,2,3}$\thanks{E-mail: jiaxin.han@sjtu.edu.cn}
Hongyu Gao$^{1,2,3}$
and Jiajun Zhang$^{4,5}$
\\
$^{1}$Department of Astronomy, School of Physics and Astronomy, Shanghai Jiao Tong University, Shanghai, 200240, People’s Republic of China\\
$^{2}$Key Laboratory for Particle Astrophysics and Cosmology (MOE), Shanghai 200240, China\\
$^{3}$Shanghai Key Laboratory for Particle Physics and Cosmology, Shanghai 200240, China\\
$^{4}$Shanghai Astronomical Observatory (SHAO), Nandan Road 80, Shanghai 200030, China\\
$^{5}$University of Chinese Academy of Sciences, Beijing 100049, China
}

\date{Accepted XXX. Received YYY; in original form ZZZ}

\pubyear{2015}

\begin{document}
\label{firstpage}
\pagerange{\pageref{firstpage}--\pageref{lastpage}}
\maketitle

% Abstract of the paper
\begin{abstract}
Using a set of high-resolution N-body simulations, we extend the unified distribution model of cold dark matter (CDM) subhaloes to the warm dark matter(WDM) case. The same model framework combining the unevolved mass function, unevolved radial distribution, and tidal stripping can predict the mass function and spatial distribution of subhaloes in both CDM and WDM simulations. The dependence of the model on the DM particle property is universally parameterized through the half-mode mass of the initial power spectrum. Compared with the CDM model, the WDM model differs most notably in two aspects. 1) In contrast to the power-law form in CDM, the unevolved subhalo mass function for WDM is scale-dependent at the low mass end due to the cut-off in the initial power spectrum. 2) WDM subhaloes are more vulnerable to tidal stripping and disruption due to their lower concentrations at accretion time. Their survival rate is also found to depend on the infall mass. Accounting for these differences, the model predicts a final WDM subhalo mass function that is also proportional to the unevolved subhalo mass function. The radial distribution of WDM subhaloes is predicted to be mass-dependent. For low mass subhaloes, the radial distribution is flatter in the inner halo and steeper in the outer halo compared to the CDM counterpart, due to the scale-dependent unevolved mass function and the enhanced tidal stripping. The code for sampling subhaloes according to our generalized model is available at \url{https://github.com/fhtouma/subgen2}.%\jx{The code for sampling subhaloes according to our generalized model is available at https://github.com/fhtouma/subgen2.} %These differences result in a mass-dependent radial distribution of WDM subhaloes which also depends on the WDM particle mass, while both dependencies are well reproduced by our model.
 
\end{abstract}


%%%%%%%%%%%%%%%%%%%%%%%%%%%%%%%%%%%%%%%%%%%%%%%%%%

%%%%%%%%%%%%%%%%% BODY OF PAPER %%%%%%%%%%%%%%%%%%

\section{Introduction}
% \jx{Suggested outline:

% subhaloes are ubiquitous and important: 
% Hierarchical structure formation leads to subhaloes. These structures determines the small scale structure. The origin and properties of subhaloes determines the properties of galaxies within. The abundance, structure of subhaloes and their connections to galaxies also provide sensitive probes for dark matter properties.

% How to model subhaloes: simulations, sam (Taylor \& Babul..), simplified sam (Jiang, vdB..), analytical/statistical model (Han16). Recently CUSP model tried to explain Han16 from first principles. 

% Han16 based on cdm. WDM known to differ (refs..). Spurious halo challenge. Analytical model missing. This work provides such a model.
% }

%Dark matter halo is a key prediction of the theory of the hierarchical structure formation. 
In the hierarchical universe, dark matter haloes form at the peaks of the initial density field and grow by accreting the surrounding matter \citep{1985ApJ...292..371D, 1986ApJ...304...15B}. During this process, low-mass haloes merge into larger ones, leading to the formation of halo substructures also known as subhaloes \citep[e.g.,][]{1998ApJ...499L...5M, 1998MNRAS.300..146G, 2005MNRAS.356.1327G, 2009ApJ...692..931L}. %According to galaxy formation theory, galaxies form within the gravitational potential of dark matter haloes. 
Galaxies embedded in a halo are carried along during a merger, leading to the existence of satellite galaxies \citep[e.g.,][]{1999ApJ...522...82K, 2001ApJ...548...33B, 2002MNRAS.335L..84S}. Understanding the evolution and distribution of dark matter subhaloes is thus crucial for studying the galaxies they host, as well as for describing the small scale distribution of dark matter which could provide sensitive probes on the nature of dark matter~\citep[e.g.,][]{2018PhR...761....1B, 2021PhRvL.126i1101N}.% Subhaloes also provide sensitive probes of dark matter properties through their abundance, structure and connections to galaxies.

The abundance and spatial distribution of subhaloes within their host halo have been extensively studied using high-resolution N-body simulations. The abundance of subhaloes is often described by the subhalo mass function, which is found to have a simple, universal form. At the low mass end, this mass function follows a power-law form with a power index of $\sim-0.9$ \citep[e.g.,][]{2004MNRAS.355..819G, 2007ApJ...667..859D, Springel08, 2010MNRAS.404..502G, 2012MNRAS.425.2169G, 2014MNRAS.438.2578G, 2016ApJ...818...10G}, close to that of the present-day halo mass function. Since accreted subhaloes cannot be larger than their host halo, an exponential cut-off is introduced at the high-mass end~\citep[e.g.,][]{2005MNRAS.359.1029V, 2008MNRAS.386.2135G, 2009MNRAS.399..983A}. The normalization of the subhalo mass function depends on the mass of the host halo, with more massive haloes containing more subhaloes. When the mass of a subhalo is described by its value at infall time, the mass function is known as the unevolved subhalo mass function~\citep{2005MNRAS.359.1029V}, which takes on a similar form as the final subhalo mass function~\citep{2018MNRAS.474..604H} but with a higher normalization. The spatial distribution of subhalo is also known to follow a universal radial profile which is independent of the subhalo mass \citep{2000ApJ...544..616G, 2004MNRAS.352..535D, 2004MNRAS.352L...1G, 2005MNRAS.363..146L, 2005ApJ...618..557N, Springel08, 2009MNRAS.399..983A}. Compared with the total density profile of the host halo, the subhalo radial profile is less centrally concentrated. %The difference between the spatial distribution of the subhalo and the smooth DM is called "anti-bias". 

Many efforts have been devoted to understanding and predicting the above population properties of subhaloes. For example, starting from the halo merger trees generated according to the extended Press-Schechter theory~\citep[][]{1993MNRAS.262..627L}, semi-analytic models (SAM) have been developed to follow the orbital and structural evolution of individual subhaloes~\citep{2001ApJ...559..716T, 2002MNRAS.333..156B, 2004MNRAS.348..811T, 2005MNRAS.364..515T, 2005MNRAS.364..535T, 2005ApJ...624..505Z, 2005MNRAS.364..977P}, accounting for various physical effects including tidal heating, tidal disruption and dynamic friction. By calibrating the model parameters with numerical simulations, SAM can evolve the subhaloes step by step and obtain their final mass, velocity, and spatial distributions. A simplified version of subhalo SAM was developed in \citet{2005MNRAS.359.1029V}. Instead of evolving individual subhaloes step by step, they utilized the average mass loss rate to model the final distribution of subhaloes \citep{2016MNRAS.458.2848J, 2016MNRAS.458.2870V, 2021MNRAS.502..621J}.

\defcitealias{Han16}{Han16}

Although these semi-analytical subhalo models can provide useful insights into various physical processes in subhalo evolution, their semi-analytic nature hinders direct interpretation of some of the resulting subhalo properties. A concise analytical model was developed in \citet{Han16}(hereafter \citetalias{Han16}),  focusing on the mass and spatial distribution of subhaloes. They showed that the spatial distribution of subhaloes approximately traces the underlying dark matter distribution if the mass of a subhalo does not evolve. Accounting for the evolution of subhalo mass introduces a radial-dependent selection function onto the subhalo radial profile. The complete model describes the joint distribution of subhaloes in their infall mass, final mass and host-centric distance, across host haloes of different masses. This model has been widely adopted in studying many theoretical and observational problems related to subhaloes, including the indirect detection of dark matter~\citep[e.g.,][]{ Hutten16}, gravitational lensing by subhaloes~\citep[e.g.,][]{Li16, Inoue16, Nierenberg17, Dai18, Wang23}, distribution and disruption of dwarf galaxies~\citep[e.g.,][]{Ferrarese16, Dooley17} and the formation of globular clusters through subhalo collisions~\citep{Madau20}. Very recently, efforts have been made to derive the assumptions of the model from first principles~\citep{SalvadorSole22a,SalvadorSole22b, SalvadorSole22c}.

Confronting these theoretical results with observations of the nearby universe, however, has led to a few serious challenges to the standard cosmological model, commonly known as the small-scale crisis~\citep[see][for a review]{2017ARA&A..55..343B}. Among these, both the ``missing satellite" problem~\citep{1999ApJ...522...82K, 1999ApJ...524L..19M} and the ``too big to fail" problem~\citep{2012MNRAS.422.1203B, 2014MNRAS.444..222G} concern the over-abundance of subhaloes produced in simulations compared to the observed population of satellites. One promising way to resolve these problems is by replacing cold dark matter (CDM) with warm dark matter (WDM) in the standard model. Due to the suppressed small scale power in the WDM universe\citep{2001ApJ...556...93B, 2005PhRvD..71f3534V}, a WDM halo harbors much fewer subhaloes than its CDM counterpart~\citep{2013MNRAS.433.1573S, 2013MNRAS.434.3337A, 2014MNRAS.439..300L, 2017MNRAS.464.4520B, 2021MNRAS.506..128B}, naturally avoiding the small scale problems. Future observations from gravitational lensing~\citep[e.g.,][]{Li16, Dai18} and stellar dynamics~\citep[e.g.,][]{2018ApJ...861...69C, 2019ApJ...880...38B} may provide more complete measurements of the spatial and mass distribution of subhaloes both with and without galaxies, providing further tests on the nature of dark matter.

To reliably and efficiently distinguish between CDM and WDM when compared with observations, it is necessary to develop an analytical model for the WDM subhalo distribution. Compared with the CDM case, there have been relatively fewer studies on the analytical properties of WDM subhaloes. This may be partly because of the existence of spurious haloes in WDM simulations~\citep{2007MNRAS.380...93W}, which contaminates the population of true subhaloes. Besides, the distribution of WDM subhaloes is also more complicated compared with their CDM counterparts. For example, \citet{2017MNRAS.464.4520B} demonstrates that the spatial distribution of WDM subhaloes is mass-dependent, making it difficult to provide a simple model for the radial profile. In addition, the concentrations of low mass haloes also differ between the CDM and WDM universes~\citep{2012MNRAS.424..684S,2016MNRAS.455..318B}, which would result in different amounts of tidal stripping for subhaloes. Some attempts have been made to extend EPS theory to WDM, but these attempts still use CDM recipes when dealing with the nonlinear evolution of subhaloes \citep{2014ApJ...792...24P}. 

In this work, we aim to extend the model developed by \citetalias{Han16} for CDM to WDM models with different DM particle masses. To this end, we carry out a series of N-body simulations for both CDM and WDM universes and extend the {\tt\string HBT+} subhalo finder to remove spurious haloes and subhaloes. According to these simulations, the \citetalias{Han16} model is modified to account for the different unevolved mass functions and different mass evolution of WDM subhaloes, and properly parametrized to account for the physical dependence on the WDM particle mass.

The structure of the paper is organized as follows. In Section~\ref{sec:data}, we present the simulation sets and subhalo catalogue used in our work. In Section~\ref{sec:framework}, we briefly review the model in the CDM case. In Section~\ref{sec:components}, we modify and calibrate the model components for WDM subhaloes. In Section~\ref{sec:results}, we compare the model predictions with simulations. Key insights on the radial distribution of WDM subhaloes are illustrated in Figure~\ref{sub-radial-demo}. Some discussions are given in Section~\ref{sec:discussion} and the summary is given in Section~\ref{sec:summary}.

%In this paper, we use $m_{\chi}$ to denote the WDM particle mass in particle physics, $m$ for the final mass of a subhalo, $m_{\mathrm{acc}}$ for the accreted (unevolved) mass of a subhalo, and $M_{$ as the virial mass of the host halo defined with a spherical density according to the top-hat spherical collapse model~\citep{Bryan&Norman98}.

\section{Data}\label{sec:data}

\subsection{The Kanli Simulations}
%\jx{many typos need to fixed. Also read the common grammar mistakes on teams}
%The purpose of our simulations is to investigate the statistical properties of warm dark matter subhaloes. 
We have performed a suite of DM-only simulations named \textit{Kanli}, including one CDM and three WDM runs. All simulations adopt the same cosmological parameters following \citet{2016A&A...594A..13P} except for different DM types, with $\Omega_m=0.3156$, $\Omega_{\Lambda}=0.6844$, $h=0.6727$, $n_s=0.967$ and $\sigma_8=0.81$. Each simulation resolves $2048^3$ particles in a cubic box of $100 h^{-1}\mathrm{Mpc}$, corresponding to an $N$-body particle mass of $1.02\times10^{7} h^{-1} M_{\odot}$, with a gravitational softening of $\epsilon=1 h^{-1}\mathrm{kpc}$. The simulations are run using the \textsc{Gadget}-4 code \citep{2021MNRAS.506.2871S}.

To obtain the initial conditions for the WDM simulation, we follow the recipe given by \citet{2001ApJ...556...93B}. The power spectrum of the WDM model can be related to the CDM model by a transfer function,
\begin{equation}
    P_{\mathrm{WDM}}(k)=T^2(k) P_{\mathrm{CDM}}(k).
\end{equation}
\citet{2001ApJ...556...93B} computed this transfer function with a full Boltzmann code, and proposed the following function to fit their results,
\begin{equation}
    T(k)=\left[1+(\theta k)^{2 \nu}\right]^{-5 / \nu}.
\end{equation}
%Here $\nu$ is a constant index, \citet{2001ApJ...556...93B} and \citet{2005PhRvD..71f3534V} found it could take a value between 1 and 1.2. \citet{2005PhRvD..71f3534V} took it as 1.12 and \citet{2014MNRAS.439..300L} adopted it as 1 for simplicity. 
Here we adopt $\nu=1.2$ as in \citet{2001ApJ...556...93B}. $\theta$ is a parameter determining the cut-off scale in the power spectrum. Its value depends on the warm dark matter particle property as
\begin{equation}
    \theta=0.05\left[\frac{m_{\chi}}{1 \mathrm{keV}}\right]^{-1.15}\left[\frac{\Omega_{\chi}}{0.4}\right]^{0.15}\left[\frac{h}{0.65}\right]^{1.3} \left[\frac{1.5}{g_{\chi}}\right]^{0.29}h^{-1} \mathrm{Mpc}.
    \label{eq:theta}
\end{equation}
Here $m_\chi$ is the thermal relic WDM particle mass, $g_{\chi}=1.5$ is the degree of freedom, and $\Omega_{\chi}$ is the WDM density parameter. The above equation may also be applied to non-thermal WDM candidates such as the sterile neutrino by converting the neutrino mass to an equivalent thermal relic particle mass~\citep{Colombi96}. 

It is convenient to define a characteristic scale known as the half-mode wavenumber, $k_{\rm{hm}}$, where the transfer function drops to $0.5$. The corresponding characteristic mass scale, the half-mode mass, $M_{\rm{hm}}$, is
\begin{equation}
    M_{\mathrm{hm}}=\frac{4 \pi}{3} \bar{\rho}\left(\frac{\pi}{k_{\mathrm{hm}}}\right)^3.
    \label{eq:Mhm}
\end{equation}
This characteristic mass scale is useful for characterizing the differences between the WDM and CDM models, which are typically found at the scale of $M_{\rm{hm}}$ or one order of magnitude higher than it \citep{2016MNRAS.455..318B}. In this paper, we choose three thermal relic WDM particle masses, $m_{\chi}=3.0, \ 1.2,\ 0.5\rm{keV}$, corresponding to the half-mode mass range of $10^{8} h^{-1} M_{\odot}$ to $10^{11}h^{-1} M_{\odot}$. They provide a large dynamic range which will help us capture our model's dependencies on $M_{\rm{hm}}$. We list the detailed parameters in Table.~\ref{Tab:Simulation Parameters}.

In principle, WDM particles should have a thermal velocity dispersion in the initial condition. However, \citet{2012MNRAS.420.2318L} suggests that the relic velocity just has negligible effects in the present day. Therefore we neglect these velocities in our initial conditions. All of our initial conditions adopt the CCVT method to create the pre-initial particle loads \citep{2018MNRAS.481.3750L}.

% Simulation Parameters table
\begin{table*}
	\centering
	\caption{The parameters related to our WDM models. $m_{\chi}$ is the thermal relic WDM particle mass. $\theta$ is the parameter defining the power spectrum cutoff, as defined in Equation~\eqref{eq:theta}. $M_{\rm{hm}}$ is the half-mode mass, as defined in Equation~\eqref{eq:Mhm}. $M_{\rm{lim}}$ is the mass scale below which the spurious haloes dominate, as defined in Equation~\eqref{eq:Mlim}. $N_{\rm{sub}}(\rm{accreted})$ is the number of all accreted subhaloes in our sample of cluster-size haloes, and  $N_{\rm{sub}}(z=0)$ is the number of subhaloes with bound particles larger than 20 at $z=0$.
        }
	\begin{tabular}{lcccccr} % four columns, alignment for each
		\hline
		$m_{\chi}/\mathrm{keV}$  & $m_{\mathrm{p}}/h^{-1}M_{\odot}$ & $\theta/h^{-1} \rm{Mpc}$ & $M_{\mathrm{hm}}/h^{-1}M_{\odot}$ & $M_{\mathrm{lim}}/h^{-1}M_{\odot}$ & $N_{\rm{sub}}(\rm{accreted})$ & $N_{\rm{sub}}(z=0)$ \\
		\hline
		CDM  & $1.02\times10^{7}$ & ~ & ~ & ~ &2291533 & 234579 \\
		0.5  & $1.02\times10^{7}$ & 0.105 & $1.1\times10^{11}$ & $8.6\times10^{9}$ &120781 & 5927\\
		1.2  & $1.02\times10^{7}$ & 0.038 & $5.4\times10^{9}$  & $1.5\times10^{9}$ &453445 & 36529 \\
            3.0  & $1.02\times10^{7}$ & 0.013 & $2.3\times10^{8}$  & $2.3\times10^{8}$ & 1284579&137563 \\
		\hline
  \end{tabular}
        
        \label{Tab:Simulation Parameters}
\end{table*}

\subsection{Subhalo catalogue}

In the \textit{Kanli} simulations, dark matter haloes are identified with the Friends-of-Friends (FoF) method \citep{1985ApJ...292..371D} with a standard linking length of $0.2$ times the mean inter-particle separation. The mass of a dark matter halo is defined as the mass in a sphere with an average density of 200 times the critical density of the universe, $M_{200}$. Starting from the halo catalogue, we use {\tt\string HBT+}~\citep{2012MNRAS.427.2437H, 2018MNRAS.474..604H} to track the evolutions of haloes and identify subhaloes as the self-bound remnants of merged haloes. Due to its unique tracking nature, {\tt\string HBT+} has been shown to have superb performance in avoiding many subhalo-finding and tree-building pitfalls common to most other subhalo-finders and merger tree builders~\citep{Muldrew11, Suss, Behroozi15, 2018MNRAS.474..604H}, resulting a highly consistent and physical catalogue of subhaloes as well as their evolutions. The mass of a subhalo, $m$, is defined to be its self-bound mass as determined by {\tt\string HBT+}. %The {\tt\string HBT+} algorithm identifies the subhalo by tracking the halo merging process and then using the binding energy to filter the subhalo, which is different from the general SubFind algorithm \citep{2001MNRAS.328..726S}. Here, the mass of the subhalo is defined as the Bound Mass.

For WDM simulations, due to the suppressed small-scale power, discrete noise will result in the production of a substantial population of spurious haloes that are not physical.
\citet{2007MNRAS.380...93W} found that these spurious haloes dominate the halo mass function below the $M_{\mathrm{lim}}$ scale, with
\begin{equation}
    M_{\mathrm{lim}}=10.1 \bar{\rho} d k_{\text {peak}}^{-2}.
    \label{eq:Mlim}
\end{equation}
Here, $d$ is the mean inter-particle separation, and $k_{\rm{peak}}$ is the wavenumber at the maximum of the dimensionless power spectrum.
To eliminate the spurious haloes and their descendant subhaloes from our catalogue, we follow \citet{2014MNRAS.439..300L} to identify them according to their shapes in the initial condition. Specifically, for each evolution track, we locate the time when the halo reaches half of its maximum mass along the history and trace the member particles at this time back to the initial condition of the simulation to define a proto-halo. The shape of this proto-halo is calculated from the eigenvalues of inertial tensor, $s=c/a$, with $c$ and $a$ being the square root of the smallest and largest eigenvalues. Spurious haloes are identified with $s<0.2$ at the low mass end. This method can remove most of the spurious haloes, although some spurious haloes in the filaments of the cosmic web are missed. These remaining spurious haloes are typically below $M_{\rm lim}$. For the $m_\chi=1.2$ and $3.0$ keV runs, the $M_{\rm lim}$ corresponds to $10-100$ particles and thus the residual spurious haloes will not contaminate the statistics of well resolved subhaloes. For $m_\chi=0.5$ keV, the $M_{\rm lim}$ is $\sim 900$ particles, and we do observe some contamination at the low mass end. We will point out such contamination explicitly when it appears, and avoid mis-interpreting them in subsequent analysis.  %which is $\sim 10-1000$ particles for our simulations. And this is equivalent to the generally believed number of well-resolved subhalo particles to meet the convergence requirements of the subhalo mass function~\citep{Springel08}. % In addition, they usually get disrupted shortly after falling into a larger halo. \textcolor{red}{jj: This sentence makes the argument suspicious. I suggest changing the way of presentation or simply removing it. In fact, a better way is to provide some quantitative proof here, some numbers, or provide some reference.}%We find that these missed spurious haloes have little impact on the final subhalo population as they usually get disrupted shortly after falling into a larger halo. %in our subsequent research, we found that many spurious subhaloes have actually been disrupted in the accreted host haloes. 
%As a result, the subhalo mass function cleaned by the empirical method is reliable enough.

In this work, we focus primarily on the analysis of subhalo samples in massive cluster-size haloes with $M_{200} = [1\sim3] \times 10^{14}h^{-1} M_{\odot}$. These massive haloes are resolved with high resolution of subhaloes comparable to those in the level-3 runs of the Aquarius project~\citep{Springel08}, allowing for a detailed investigation of subhalo distribution. There are $\sim 20$ such haloes in each simulation. The four simulations in \textit{Kanli} adopt the same random phases when generating the initial conditions whenever possible, so that these massive haloes can be matched across simulations at early time. The numbers of subhaloes in each simulation used in our samples are listed in Table.~\ref{Tab:Simulation Parameters}.

\section{The unified distribution model for CDM subhaloes}\label{sec:framework}

Based on the high resolution zoom-in simulations from the Aquarius~\citep{Springel08} and the Phoenix~\citep{2012MNRAS.425.2169G} projects, \citetalias{Han16} has developed and calibrated a unified model to describe the spatial and mass distribution of CDM subhaloes. The model is composed of the following three components:

\begin{enumerate}
\item \textbf{Unevolved mass function}: The infall (unevolved) subhalo mass function follows a universal single power-law form,
\begin{equation}
\frac{\mathrm{d} N}{\mathrm{~d} \ln m_{\mathrm{acc}}} \propto m_{\mathrm{acc}}^{-\alpha}.
\end{equation}
This relation holds well at the low-mass end, with a power-index $-\alpha=-0.9 \sim -1$. Here, the infall mass $m_{\rm{acc}}$ is defined as the maximum mass the subhalo ever had before accreted into the current host halo.

\item \textbf{Unevolved radial profile}: For subhaloes of a given infall mass, the number density profile follows the density profile of their host halo,
\begin{equation}
\frac{\mathrm{d} N\left(R \mid m_{\mathrm{acc}}\right)}{\mathrm{d}^{3} R} \propto \rho(R).
\end{equation}
Ignoring dynamical friction which is significant only for massive subhaloes at small radii, the orbits of subhaloes are expected to resemble the orbits of dark matter particles. As the density profile is a snapshot of the orbits~\citep{oPDF}, the spatial distribution of subhaloes is thus expected to follow that of dark matter particles in the same halo.

\item \textbf{Tidal stripping law}: The amount of mass stripping on a subhalo, $\mu=m/m_{\rm{acc}}$, depends on its halo-centric distance, with a median relation of
\begin{equation}
\bar{\mu}\left( R \right) = \mu_{*} \left(R / R_{200} \right)^{\beta}.\label{eq:medstrip}
\end{equation}
\end{enumerate}
Here, $\mu_{*}$ is a normalization factor and $R_{200}$ is the host viral radius.

Combining the three components above,  the final distribution of subhaloes can be derived analytically as
\begin{equation}
\frac{\mathrm{d}N\left(m,R \right)}{\mathrm{d}\ln m\mathrm{d}^3R}\propto \left[\frac{m}{\bar{\mu}(R)}\right]^{-\alpha} \rho(R).
\label{eq: CDM distribtuion}
\end{equation}
Where $\bar{\mu}$ is given by Equation~\eqref{eq:medstrip}.

Equation~\eqref{eq: CDM distribtuion} indicates that the final subhalo mass function shares the same formula as the unevolved one, and the spatial distribution depends on the halo density profile and the tidal stripping efficiency, i.e. the power-index $\gamma=\alpha\beta$. 

The complex orbital and structural distributions of subhaloes introduce a lognormal scatter around the median stripping law in Equation~\eqref{eq:medstrip}. In addition, a certain fraction of subhaloes are found to have been completely disrupted in \citetalias{Han16}. A complete model for the subhalo mass stripping is thus specified by the distribution function
\begin{equation}
\begin{aligned}
\mathrm{d} P\left(m \mid m_{\mathrm{acc}}, R\right)= & \left(1-f_{\mathrm{s}}\right) \delta(m) \mathrm{d} m \\ 
& +f_{\mathrm{s}} \mathcal{N}\left(\ln \frac{m}{m_{\mathrm{acc}}}, \ln \bar{\mu}(R), \sigma\right) \mathrm{d} \ln m . \label{eq:strip_pdf}
\end{aligned}
\end{equation}
Where $f_{\mathrm{s}}$ is the number fraction of the surviving population. Replacing Equation~\eqref{eq:medstrip} with Equation~\eqref{eq:strip_pdf} leads to a statistical model for the joint distribution of the subhalo in infall mass, final mass and halo-centric distance, $P(m, m_{\rm acc}, R)$.

The CDM subhalo distribution model has been verified in the Aquarius and Phoenix simulations. In the following, we will modify and calibrate the model components for our WDM simulations.

\section{Modification to model components in WDM}\label{sec:components}

\subsection{Unevolved Radial Number Density Profile}
\begin{figure*}
    \centering
        \includegraphics[width=\columnwidth]{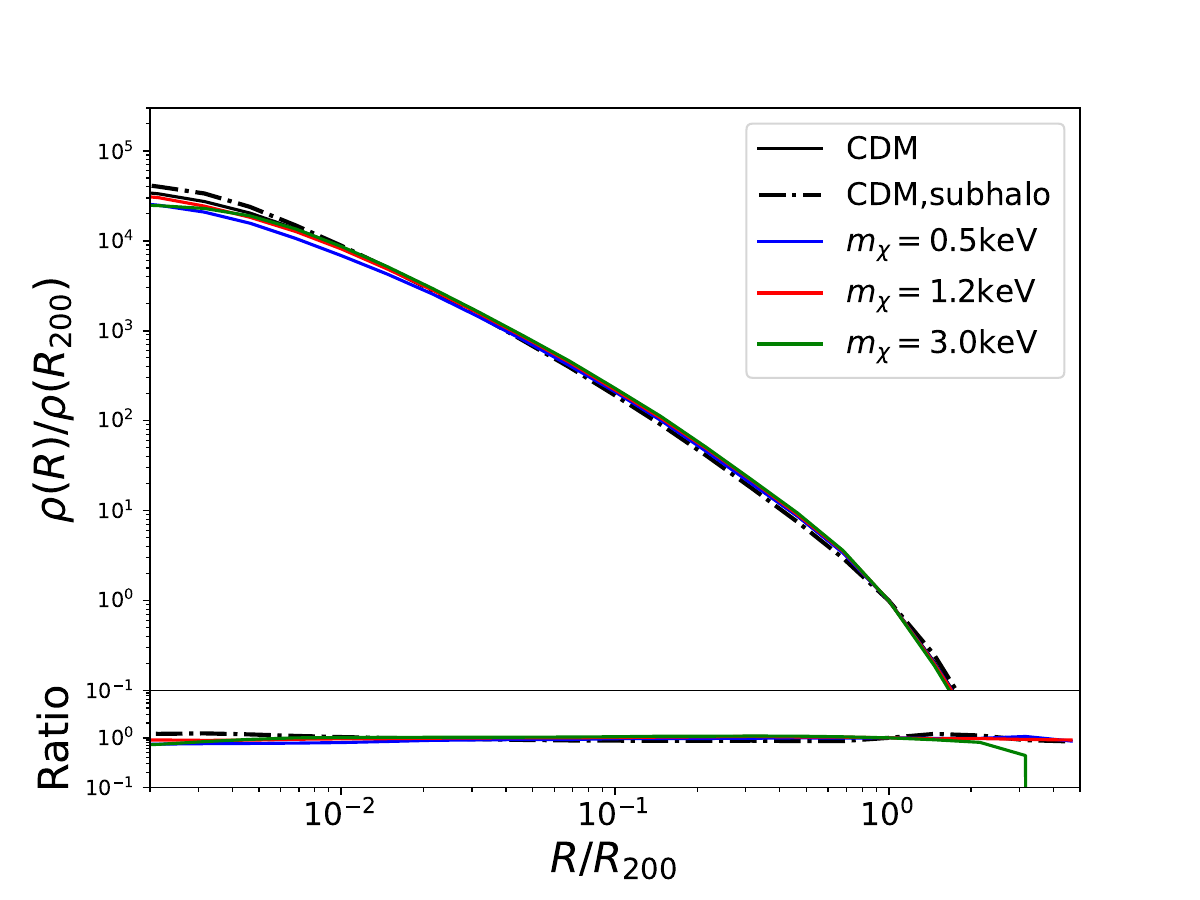}
        \includegraphics[width=\columnwidth]{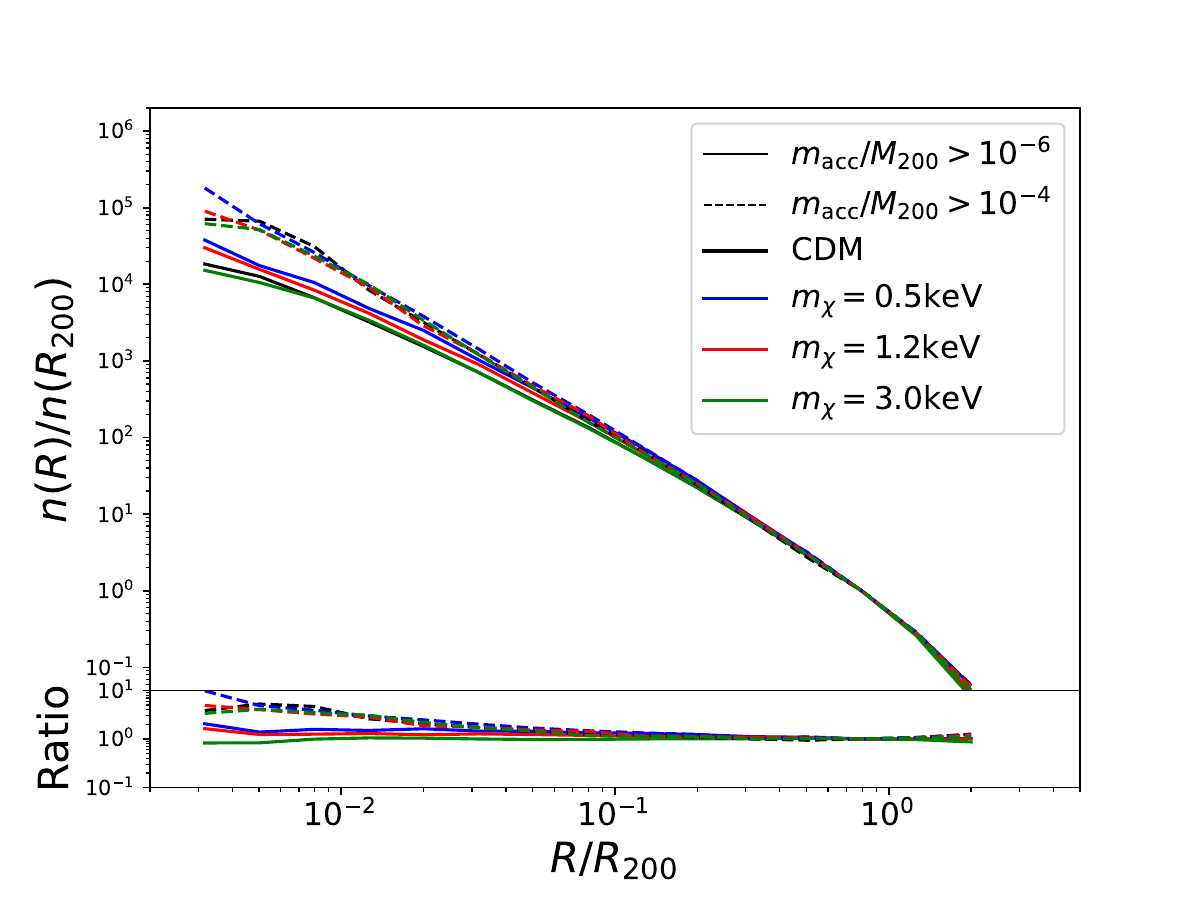}
    %}
    \caption{Left: Halo density profiles of different DM models. The solid lines of different colours correspond to the density profiles in different DM models as labelled. All of them have been normalized by the density value at $R_{200}$. The dash-dotted line corresponds to the number density profile of CDM subhaloes with $m_{\mathrm{acc}}/M_{200} > 10^{-6}$. The bottom gives the halo density ratio between different WDM models and the CDM model. The density profiles of different DM-type haloes are very close to each other. Right: Unevolved radial number density profiles of different DM-type subhaloes. The solid lines correspond to subhaloes with $m_{\mathrm{acc}}/M_{200} > 10^{-6}$, and the dashed lines correspond to subhaloes with $m_{\mathrm{acc}}/M_{200} > 10^{-4}$. The bottom panel shows the number density ratio between the different cases and the CDM case with $m_{\mathrm{acc}}/M_{200}>10^{-6}$. The higher number density for high mass subhaloes in the inner region can be attributed to dynamical friction. The unevolved number density profiles of different DM-type subhaloes are very close to each other.} %\textcolor{red}{jj: should be dynamical friction rather than dynamical fraction, right? this caption is long and confusing. A one-sentence comment for each panel can be provided, explaining the most important information we would like to show to the readers.} }
    \label{fig: density profile and number density}
\end{figure*}

The left panel of Fig.~\ref{fig: density profile and number density} shows the average mass density profile in each simulation for cluster-size haloes with $M_{200} = [1\sim3] \times 10^{14} h^{-1} M_{\odot}$. Comparing the different runs of \textit{Kanli}, the density profiles for these massive haloes are very close to each other. This is consistent with the findings in the CoCo-WDM simulation \citep{2017MNRAS.464.4520B}. Due to the suppression of the power spectrum on small scales, the primary distinction between WDM and CDM arises in small haloes while it is not markedly discernible in the cluster-size halo. 

%of cold dark matter and warm dark matter haloes.
The unevolved radial number density profile of CDM subhaloes is also shown in Fig.~\ref{fig: density profile and number density} for subhaloes with $m_{\rm acc}/M_{200}>10^{-6}$, which closely traces the density profile of the host halo and confirms the hypothesis of \citetalias{Han16}. 

The right panel of Fig.~\ref{fig: density profile and number density} shows the unevolved number density profiles for CDM and WDM subhaloes, that is, the final radial distribution of subhaloes selected according to their infall mass. Note that disrupted subhaloes are also included according to their positions traced by their most-bound particles. Overall, the profiles from different DM models are close to each other, especially in the outskirts. Only in the inner region at $R/R_{200} < 0.1$, the WDM subhalo has a slightly higher radial number density, with the largest difference observed for the lightest WDM particles. The profiles for subhaloes with $m_{\rm {acc}}/M_{200}>10^{-6}$ are also very close to the total DM density profile, while those with a larger mass cut of $m_{\rm{acc}}/M_{200}>10^{-4}$ shows a steeper inner profile, reflecting stronger dynamical frictions on these objects, in accordance with the findings of \citetalias{Han16}. These results confirm that we can still approximate the unevolved radial distribution of WDM subhaloes as following their host density profile.% In general, our model assumptions still hold approximately. For WDM with different masses, the infall radial number density profile is approximately the same as the halo density profile, that is, it conforms to the NFW profile.

\subsection{Subhalo Mass Function}

\begin{figure}
    \includegraphics[width=\columnwidth]{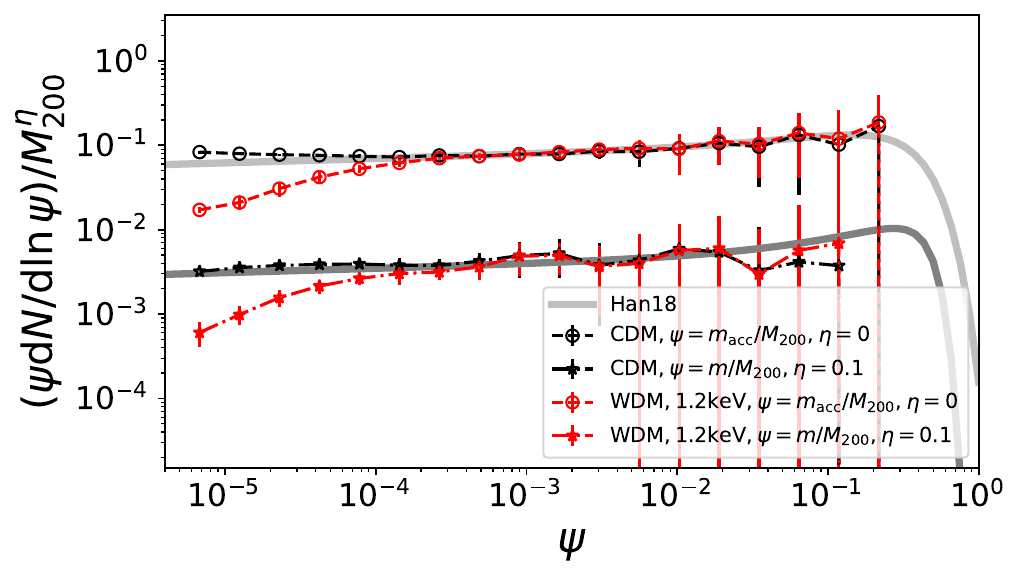}
    \caption{The unevolved and final subhalo mass functions of the CDM model and the WDM model with $m_{\chi}=1.2\mathrm{keV}$. The black lines correspond to the CDM model and the red lines to the WDM model. The top open circles mark the unevolved subhalo mass functions and the bottom solid stars correspond to the final subhalo mass functions. The grey solid lines are the best-fit subhalo mass functions given by \citet{2018MNRAS.474..604H}. }
    \label{fig:CDM SMF}
\end{figure}

In the CDM case, both the unevolved and the final subhalo mass functions can be described by a universal power-law function with a power index between - 0.9 and - 1. By analysing the tidal stripping on the subhaloes, \citetalias{Han16} connects the unevolved subhalo mass function to the final one. Fig.~\ref{fig:CDM SMF} shows the subhalo mass function in our CDM simulation. The thick grey lines represent the formula given by \citet{2018MNRAS.474..604H},
\begin{equation}
    \frac{\mathrm{d} N}{\mathrm{d} \ln \psi}=\left(\frac{M_{200}}{m_0}\right)^\eta\left(\mathcal{A}_1 \psi^{-\alpha_1}+\mathcal{A}_2 \psi^{-\alpha_2}\right) \exp \left(-\mathcal{B} \psi^\tau\right).    
\end{equation}
Where $M_{200}$ is the mass of the host halo, $m_0=10^{10} \mathrm{M}_{\odot} h^{-1}$ and $\psi=m/M_{200}$ is the mass ratio between the subhalo and its host halo. We adopt the model parameters given by \citet{2018MNRAS.474..604H}, with $\mathcal{A}_1=0.11, \alpha_1=0.95, \mathcal{A}_2=0.2, \alpha_2=0.3, \mathcal{B}=7.6, \tau=2.1$ and $\eta=0$ for the unevolved mass function, and $\mathcal{A}_1=0.0055, \alpha_1=0.95, \mathcal{A}_2=0.017, \alpha_2=0.24, \mathcal{B}=24, \tau=4.2$ and $\eta=0.1$ for the final mass function. At the low mass end, the unevolved mass function for the CDM model can be simplified to a single power-law form,
\begin{equation}
    \frac{\mathrm{d} N}{\mathrm{~d} \ln m_{\mathrm{acc}}}=A_{\mathrm{acc}} \frac{M_{200}}{m_0}\left(\frac{m_{\mathrm{acc}}}{m_0}\right)^{-\alpha}.
\end{equation}
Here $A_{\mathrm{acc}}=0.11(M_{200}/m_0)^{-0.05}$ and $\alpha=0.95$. The subhalo mass function for the WDM model with $m_{\chi}=1.2\mathrm{keV}$ is also presented in Fig.~\ref{fig:CDM SMF}. It is clearly shown that the WDM and CDM mass functions are in good agreement with each other at the high mass end, while for low mass bins, the unevolved and final mass functions of the WDM model both show a rapid drop. \citet{2012MNRAS.424..684S} and \citet{2020ApJ...897..147L} gave a fitting function for the ratio of the mass function of WDM to that of CDM haloes, analogous to the ratio between their power spectrum, as,
\begin{equation}
    n_{\mathrm{wdm}}/n_{\mathrm{cdm}} = \left(1+\left(a M_{\mathrm{hm}}/m\right)^{b}\right)^{c}.
    \label{eq: suppression function}
\end{equation}
%where %$n_{\mathrm{wdm}}$ and $n_{\mathrm{cdm}}$ are the (sub)halo mass function of WDM and CDM, 
%$M_{\mathrm{hm}}$ is the half mode mass defined in Eq.~\ref{eq: Mhm}. 
%, which means that the mass scale suppressed to half of the initial CDM power spectrum, 
Our best-fitting parameters are $a=2.3$, $b=1$ and $c=-0.68$, which work for both the unevolved and final subhalo mass functions. 

\begin{figure*}
 \includegraphics[scale=0.8]{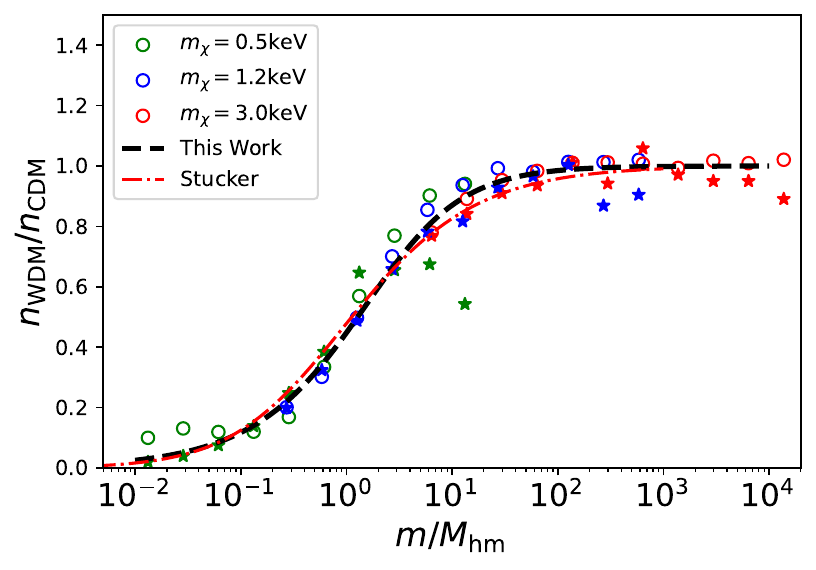}
 \caption{The ratio of the subhalo mass function between WDM model and CDM model for subhaloes in the cluster-size halo. The open circles show the accreted (unevolved) subhalo mass function and the solid stars show the final subhalo mass function. Red, blue and green markers correspond to the WDM models with $m_{\chi}=3.0\mathrm{keV}$, $1.2\mathrm{keV}$ and $0.5\mathrm{keV}$ respectively. The x-axis is the bound mass of subhaloes scaled by each half-mode mass of the WDM model. The black dashed line gives the fitting result of Equation~\eqref{eq: suppression function}, and the red dash-dotted line gives the results from \citet{2022MNRAS.509.1703S}. The unevolved subhalo mass function of $0.5\mathrm{keV}$ WDM is slightly higher than the expectation due to some residual spurious subhaloes. %At the high mass end, few subhalo numbers make the higher statistical error. 
 The unevolved and final subhalo mass functions exhibit the same functional forms.} 
 \label{fig:subhalo mass function suppression}
\end{figure*}

Fig.\ref{fig:subhalo mass function suppression} shows the subhalo mass function ratio between WDM and CDM in cluster-size haloes of the \textit{Kanli} simulation. The results given by \cite{2022MNRAS.509.1703S} are also shown. %The average mass of the host halo mass of this subhalo sample is $2\times10^{14} h^{-1} M_{\odot} $, and the horizontal axis is the mass scaled by $M_{\mathrm{hm}}$. 
It should be pointed out that the result of \citet{2022MNRAS.509.1703S} is the global mass function for all subhaloes in the simulation box, whereas our result is for subhaloes in a host halo with a given mass. In addition, due to the difference in the halo finder, the final mass functions are slightly different.

\begin{figure}
    \includegraphics[width=\columnwidth]{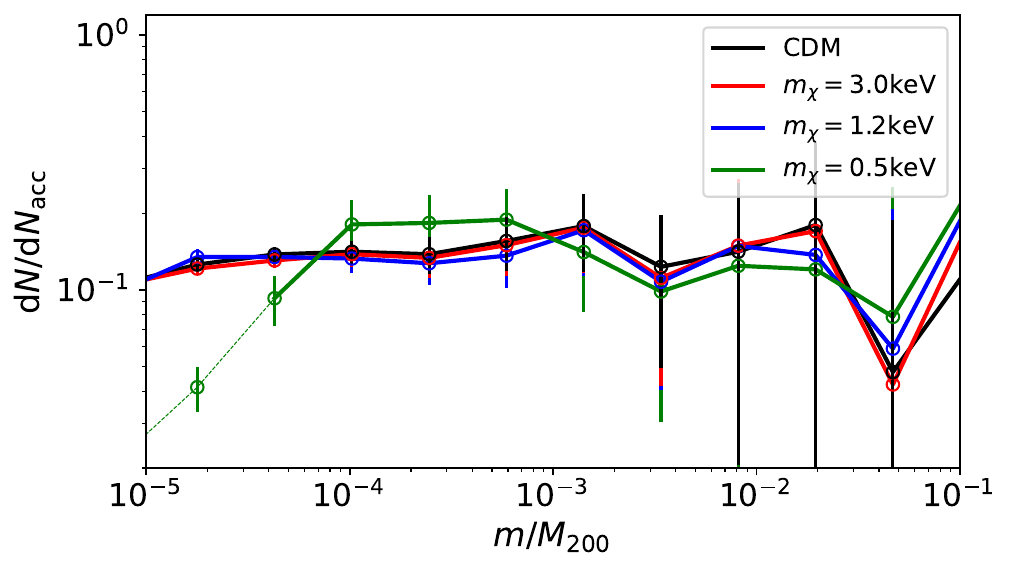}
    \caption{The ratio between the unevolved and final subhalo halo mass functions with Poisson error bars. The different colours correspond to different DM models as labelled. The mass-independent ratio between the unevolved and final mass functions of the CDM subhalo is expected from \citetalias{Han16}. The ratios for the WDM subhaloes follow the same behaviour. The drop-off in $m_{\chi}=0.5\mathrm{keV}$ case due to numerical effects is denoted by the thin green line.} %\textcolor{red}{jj: For this figure, I think we may need to provide some error bars for the line, in order to clarify whether the difference is beyond statistical error. A Poisson error estimation may be good enough.}}
    \label{fig:dNdNacc}
\end{figure}

It can be seen that for WDM, both the unevolved and the final subhalo mass functions follow approximately the same functional form. This is in line with the conclusion in the CDM case. Fig.~\ref{fig:dNdNacc} further verifies this conclusion, showing the ratio between the final mass function and the unevolved mass function for the CDM model and the WDM models with different particle masses, respectively. In the CDM case, the final mass function can be broadly understood as a horizontally shifted version of the unevolved mass function due to an average mass loss ratio that is independent on the subhalo mass. However, for WDM, a simple horizontal shift would result in a non-constant ratio between the unevolved and final mass functions. This is because the unevolved mass function is no longer scale-free but declines around the $M_{\rm hm}$ scale. We will see below that the constant ratio in the WDM case is further contributed by the mass-dependent stripping law of WDM subhaloes.

\subsection{Mass Stripping}

\begin{figure*}
    \centering
    %\subfigure{
        \includegraphics[width=\columnwidth]{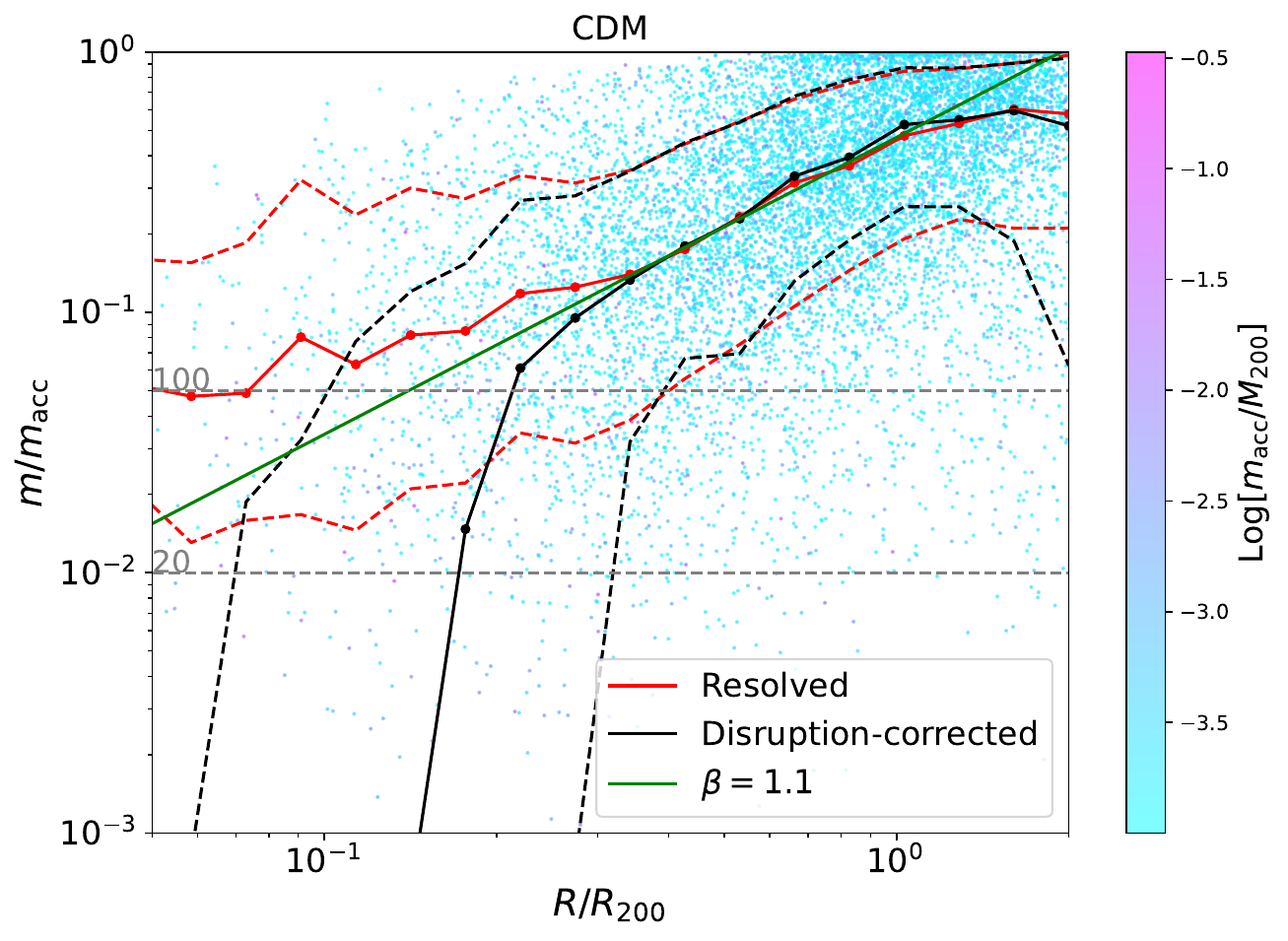}
    %}
    %\subfigure{
        \includegraphics[width=\columnwidth]{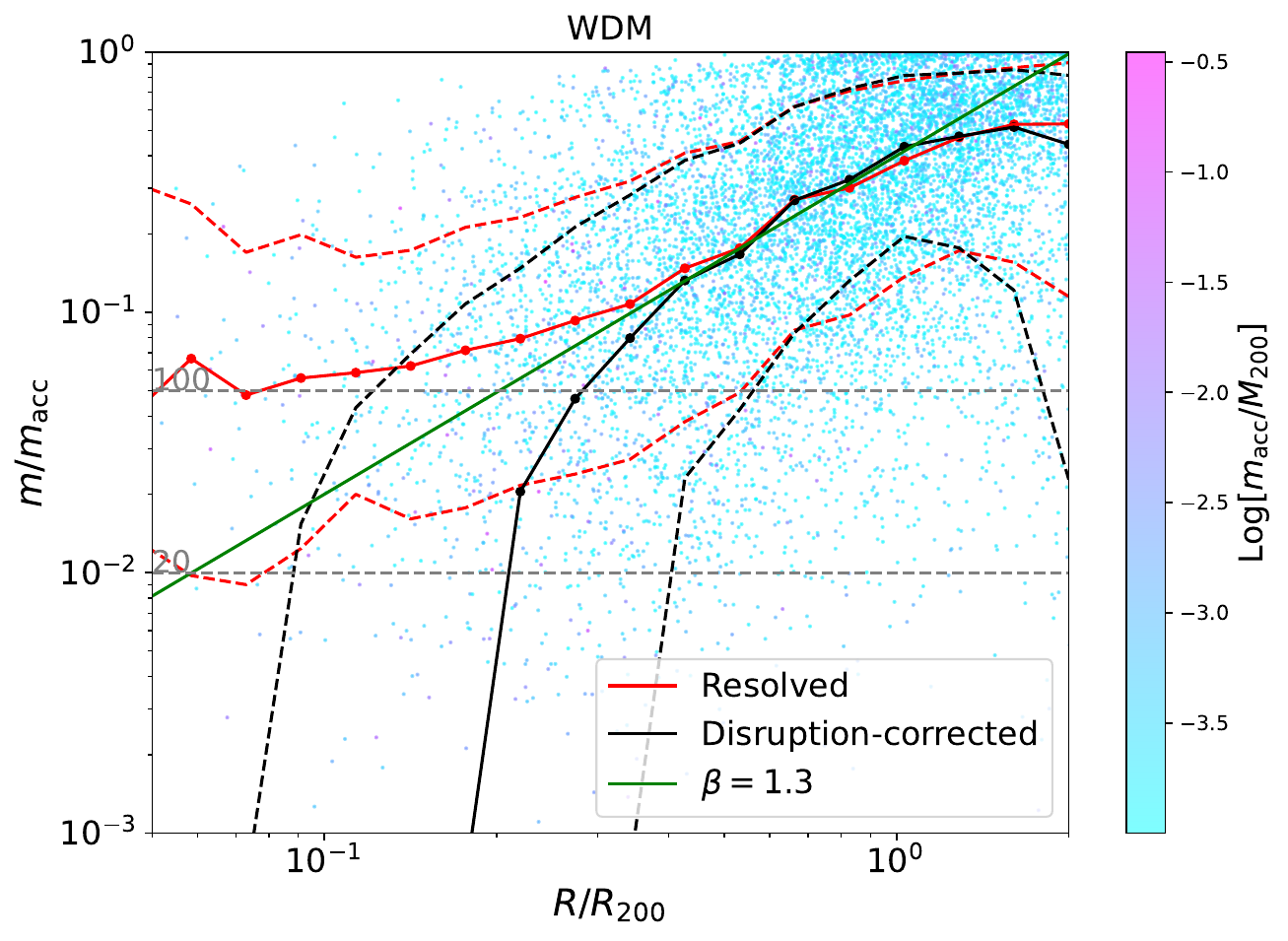}
    %}
    \caption{The stacked distribution of subhalo bound fraction versus halo-centric radius. The left and right panels show the results for CDM and WDM with $m_{\chi}=1.2\mathrm{keV}$, respectively. The blue dots in the background show the measurements for individual subhaloes. The colour bar on the right of the figures marks the logarithm of the subhalo infall mass. All subhaloes in the cluster-size halo samples with $m_{\mathrm{acc}}/M_{200} > 10^{-4}$ have been stacked in this figure. The red solid line corresponds to the median value of the bound fraction of resolved subhaloes at each radius, while the two red dashed lines give the 16th and the 84th percentiles at each radius. The corresponding black lines show the result when both resolved and disrupted subhaloes are considered (see text for detail). The green solid line is the fitted median of the bound fraction. The grey dashed lines correspond to the completeness limit, corresponding to subhaloes with more than 20 and 100 bound particles at the present time. The tidal stripping in the WDM case is much stronger than that in the CDM case.}
    \label{fig:Mass Stripping}
\end{figure*}

\begin{figure}
    \includegraphics[width=\columnwidth]{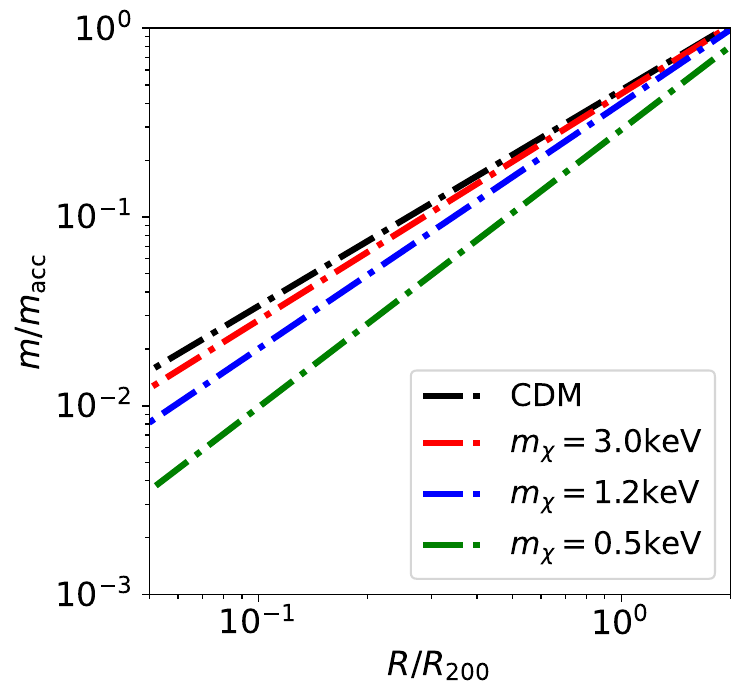}
    \caption{The median mass stripping model for subhaloes in CDM and three WDM cases.}
    \label{fig: median mass stripping}
\end{figure}

As a subhalo orbit in the host halo, it can undergo significant mass loss due to tidal heating and tidal stripping. The detailed mass loss process depends on many factors including the orbit, the infall time, and the internal structures of the subhalo and the host. \citetalias{Han16} found that the amount of mass stripping can be described statistically by Equation~\eqref{eq:strip_pdf}, which gives the probability of the final mass distribution of the subhalo for a given infall mass and halo-centric radius. The first term of the equation describes fully disrupted subhaloes, while the second term describes the distribution of the surviving ones. We first examine the second term for WDM subhaloes, and leave the first term for further discussion in the next subsection.

Fig.~\ref{fig:Mass Stripping} shows the stacked subhalo bound fraction distribution in the CDM haloes and $1.2\mathrm{keV}$ WDM haloes. Only resolved subhaloes at the present day are shown. It is likely that the subhaloes with a small number of particles are poorly or partly resolved. For reference, we plot two completeness limits (grey dashed lines) above which all the subhaloes are resolved with 20 and 100 particles respectively. The red lines represent the median (solid line) and the $\pm 1 \sigma$ percentiles (dashed lines) of the bound fraction distribution at each radial bin. Because unresolved subhaloes would contribute to the lower part of this plot, the red curves should be interpreted as upper limits to the actual percentiles for the surviving subhaloes. 

To complement this, we also show another set of curves representing lower limits to the actual percentiles following~\citetalias{Han16}. According to the mass distribution model of Equation~\eqref{eq:strip_pdf}, the percentiles of the surviving population can also be estimated from percentiles of the total population including disrupted ones. Specifically, the $p$-th  percentile of the surviving population would correspond to a $p^{\prime}$-th percentile of the full sample, with $p^{\prime} / 100=\left(1-f_{\mathrm{s}}\right)+f_{\mathrm{s}} p / 100 $. Even though we do not include unresolved or disrupted subhaloes in Fig.~\ref{fig:Mass Stripping}, we do know their numbers and expect them to contribute to the lower part of the plot. As long as the $m/m_{\rm acc}$ of these unresolved subhaloes lie below the percentiles we are interested in, we can safely extract those percentiles. When some unresolved subhaloes have $m/m_{\rm acc}$ above the percentile of interest, the extracted percentile becomes a lower limit. The black curves show the corresponding percentiles extracted this way.

In the outer halo, the two sets of percentiles overlap with each other, converging to the true percentiles. In the inner halo, the two sets diverge, and the true percentiles are expected to lie in between the two. A power-law fit (Equation~\eqref{eq:medstrip}) to the converge part of the bound fraction yields $\beta=1.1$ for CDM subhaloes. This is consistent with the $\beta=1.0$ found by \citetalias{Han16} in the Phoenix simulations. Note our results are based on {\tt\string HBT+}, while \textsc{Subfind} was used in \citetalias{Han16}. 

%Clearly, in the CDM case, the median bound fraction of subhaloes follows a power-law relationship with the radius. At each radius bin, the bound fraction follows a log-normal distribution, as expressed in Eq.~\ref{eq:mixed distribution}. Meanwhile, the subhaoes in the WDM case have a similar rule to the CDM case. 
The right panel of Fig.\ref{fig:Mass Stripping} shows the result of the WDM model with $m_{\chi}=1.2\mathrm{keV}$. Overall, the distribution of the mass fraction is similar to that for the CDM subhaloes, while quantitatively the median relation is steeper, with $\beta=1.3$. Fig.\ref{fig: median mass stripping} compares the fitted median mass stripping functions for subhaloes in the CDM and three WDM models. It is clear that tidal stripping is stronger for WDM models with a smaller $m_\chi$. This can be understood as WDM subhaloes with a smaller $m_\chi$ tend to be less concentrated at the accretion time~\citep{2012MNRAS.424..684S, 2014ApJ...792...24P, 2016MNRAS.455..318B}, making them more susceptible to tidal stripping. %In the WDM case, we observe stronger tidal stripping of the subhaloes, which results in a higher power index $\beta$ than that in the CDM case. While the number of WDM subhaloes is one or two orders of magnitude less than that of CDM subhaloes, this statistical tendency is still apparent. It is worth noting that the tidal stripping efficiency $\beta$ becomes higher as the WDM particle mass decreases. The power suppression in the $0.5\mathrm{keV}$ case is considerably greater than that in the $3.0\mathrm{keV}$ and $1.2\mathrm{keV}$ cases, resulting in markedly distinct properties of subhaloes.

% Model Parameters table
\begin{table*}
	\centering
	\caption{The model parameters in Equations~\eqref{eq:medstrip} and \eqref{eq:strip_pdf} describing the stripped mass distribution, extracted from the cluster subhalo sample. The first row shows the parameters in the CDM model, and the next three rows show the parameters of the three WDM models. The final row gives a generalization of the parameters to arbitrary $M_{\rm{hm}}$, with $m_0\equiv10^{10}h^{-1}M_{\odot}$. 
 } 
	\label{tab:model_parameters_table}
	\begin{tabular}{lcccr} % four columns, alignment for each
		\hline
		$m_{\chi}/\mathrm{keV}$ & $M_{\mathrm{hm}}/h^{-1}M_{\odot}$ & $\mu_{*}$ & $\beta$ & $\sigma$ \\ 
		\hline
		CDM & 0             & 0.48 & 1.12 & 0.99 \\
            3.0 & $2.3\times10^{8}$  & 0.46 & 1.2  & 1.01 \\
            1.2 & $5.4\times10^{9}$  & 0.42 & 1.3  & 1.1 \\
		0.5 & $1.1\times10^{11}$ & 0.29 & 1.45 & 1.25   \\
  \hline
            General & $M_{\rm hm}$ & $\mu_*^{\rm CDM}(1 + (M_{\mathrm{hm}}/m_0) ^{0.55})^{-0.3}$ & $\beta^{\rm CDM}(1 + (M_{\mathrm{hm}} / m_0)^{0.25})^{0.27}$ & $\sigma^{\rm CDM}(1 + (M_{\mathrm{hm}} / m_0)^{0.44})^{0.17}$ \\
		\hline
	\end{tabular}
\end{table*}

We list the parameters governing the stripped mass distribution in Table~\ref{tab:model_parameters_table}. Note the parameters for the CDM subhaloes are close to but not the same as those found by \citetalias{Han16} from the Phoenix simulations. The largest difference is found for the $\mu_*$ parameter, with $\mu_*=0.34$ according to \citetalias{Han16}. The difference can be attributed to the different subhalo finders used, which is \textsc{Subfind} in \citetalias{Han16} but \textsc{HBT+} here. As shown in \citet{2018MNRAS.474..604H}, \textsc{HBT+} tends to return a systematically smaller difference between the unevolved and evolved subhalo mass functions due to its robustness in tracking the subhalo evolution. 

To incorporate the dependencies on the DM particle property into a general model, we relate them to the $M_{\rm{hm}}$ parameter by the following equation,
\begin{equation}
    X_{\rm{WDM}}/X_{\rm{CDM}}=\left(1 + \left(M_{\rm{hm}}/m_0 \right)^{p_1} \right)^{p_2}.
    \label{eq:parameter_Mhm}
\end{equation}
Here, $X_{\rm{CDM}}$ is the parameter value in the CDM model and $X_{\rm{WDM}}$ is that in the WDM model. $p_1$ and $p_2$ are fitting parameters. The reason we chose this form of Equation~\eqref{eq:parameter_Mhm} is that for a small $M_{\rm{hm}}$ corresponding to a large $m_{\chi}$, the subhalo population approach the CDM population. On the other hand, when $m_{\chi}$ is small and $M_{\rm{hm}}$ is large, the WDM model parameters will significantly deviate from that of the CDM model. The fitted results using Equation~\eqref{eq:parameter_Mhm} are listed in the bottom row of Table~\ref{tab:model_parameters_table}.

\subsection{Survival Rate}\label{sec:survival}

\begin{figure*}
    \centering
    %\subfigure{
        \includegraphics[width=\columnwidth]{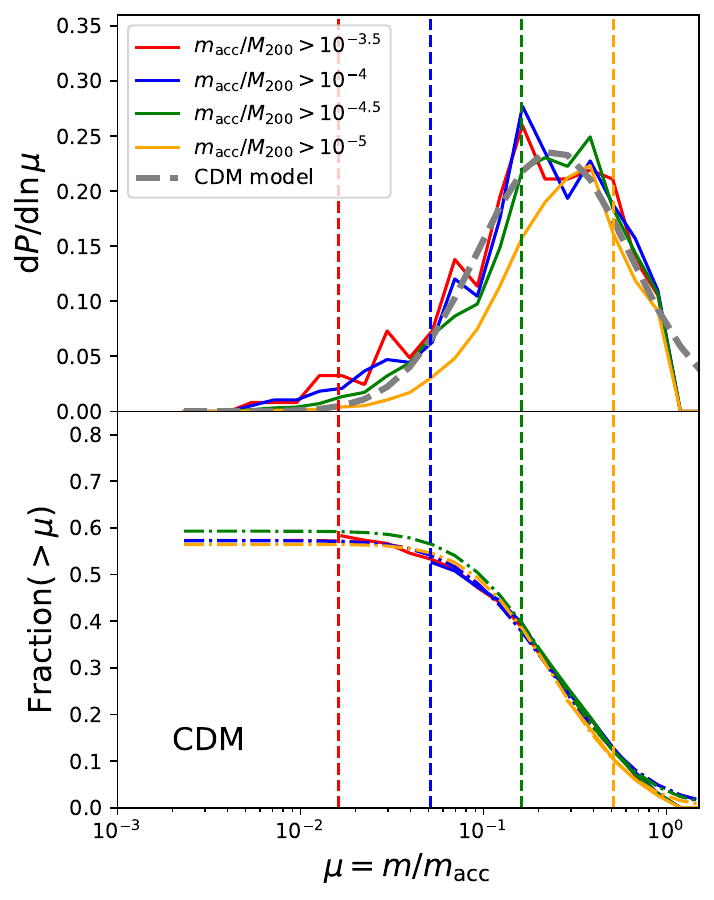}
    %}
    %\subfigure{
        \includegraphics[width=\columnwidth]{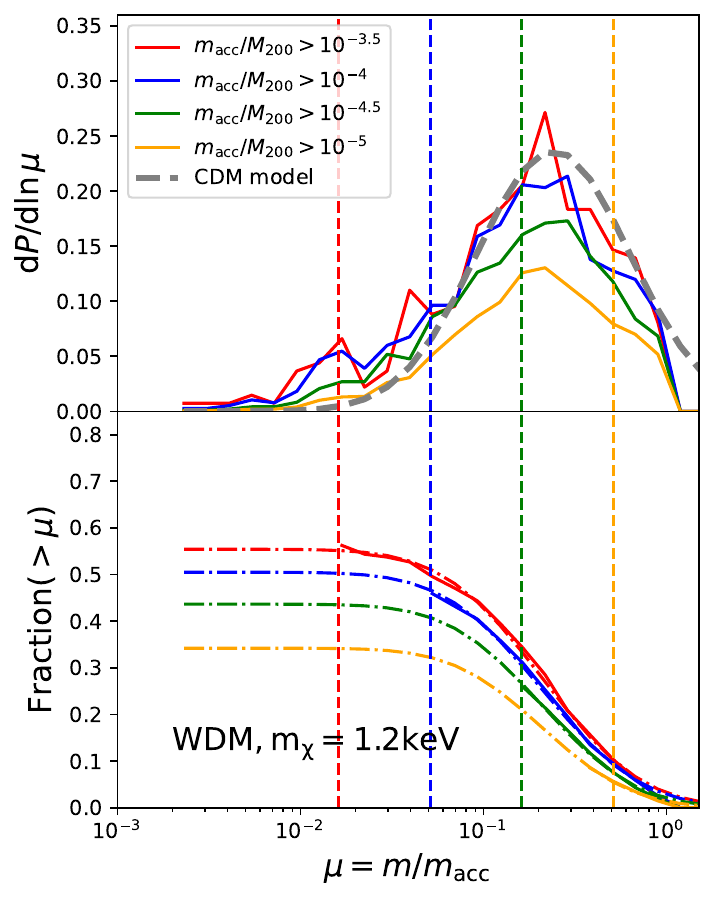}
    %}

    \caption{Top: the probability distribution of $\mu=m/m_{\mathrm{acc}}$ for subhaloes in the radial range of $0.45 \sim 0.55R/R_{200}$  with different infall masses. The grey dashed line corresponds to the log-normal distribution model used in our CDM model. The vertical dashed lines correspond to the completeness limit (at 100 bound particles) for each infall mass bin. Bottom: The solid lines show the cumulative distribution of $\mu$ for the different infall mass bins. The dash-dotted lines correspond to the extrapolated cumulative distribution of $\mu$ according to a log-normal distribution. The left column shows the results in the CDM case, and the right column shows the results in the WDM model with $m_{\chi} = 1.2\mathrm{keV}$.}
    \label{fig:lognormal }
\end{figure*}

Fig.\ref{fig:lognormal } shows the detailed distribution of the subhalo bound fraction in the CDM and the $1.2\mathrm{keV}$ WDM models. %According to Eq.~\ref{eq:average mass stripping}, this distribution is radial dependent. In order to take more subhalo samples into account, we present the probability distribution of $\tilde{\mu}$ rather than $\mu$, where $\tilde{\mu}=(m/m_{\mathrm{acc}})/(R/R_{200})^{\beta}$. 

According to the model of \citetalias{Han16}, around $45\%$ of CDM subhaloes ever accreted will become disrupted at $z=0$, independent of subhalo infall masses, while the remaining ones survive to the present day. %, and this proportion will not change with subhalo infall mass.  
In the previous subsection, we found that WDM subhaloes are more susceptible to tidal effects than CDM subhaloes, so it is expected that the survival fraction of WDM subhaloes should be lower than that of CDM subhaloes. In this subsection, we will estimate this fraction from our simulation data.

In principle, one could directly count the number of disrupted and surviving subhaloes to get their fractions. However, for any simulation with a finite resolution, the surviving population can only be resolved at most down to a certain mass limit, %(20 particles in our case), 
while those below the mass limit will be recorded with 0 mass and become degenerate with physically disrupted ones. This prevents us from directly counting their contributions. %in the numerical simulation, the disrupted subhalo we obtained includes the physically disrupted population and the unresolved population. Therefore, the unresolved population will affect our ability to obtain the truly disrupted fraction. In contrast, we investigate the resolved population to get the true fraction. As we have mentioned in the last section, the resolved subhaloes suffer from the completeness limit. 

To overcome this problem, we will only use a complete population of well-resolved subhaloes to infer the model parameters. In Fig.~\ref{fig:lognormal } we show the distributions of $\mu$ for subhaloes of different infall masses. For each infall mass limit, the vertical dashed line of the same colour shows the completeness limit, $\mu_{\rm lim}=m_{\rm lim}/M_{\rm 200,min}$, where $m_{\rm lim}$ equals 100 particle mass and $M_{\rm 200,min}$ is the minimum host halo mass in the subsample. Above this limit, all the subhaloes are well resolved with more than 100 bound particles, while the sample is incomplete below it.

As shown in the left panel of Fig.~\ref{fig:lognormal }, the distribution of the bound fraction for the complete population agrees well with the lognormal distribution model. As a result, we can fit this part of the distribution to obtain the model parameters in Equation~\eqref{eq:strip_pdf}, including the survival fraction parameter $f_s$. This parameter can also be directly obtained from the asymptotic value of the cumulative distribution at the low $\mu$ end, resulting in $f_s\simeq 0.58$, consistent with the result of \citetalias{Han16}. The cumulative distributions with different infall masses converge well within their corresponding completeness limits, consistent with a universal $\mu$ distribution independent of infall mass.

\begin{figure}
    \includegraphics[width=\columnwidth]{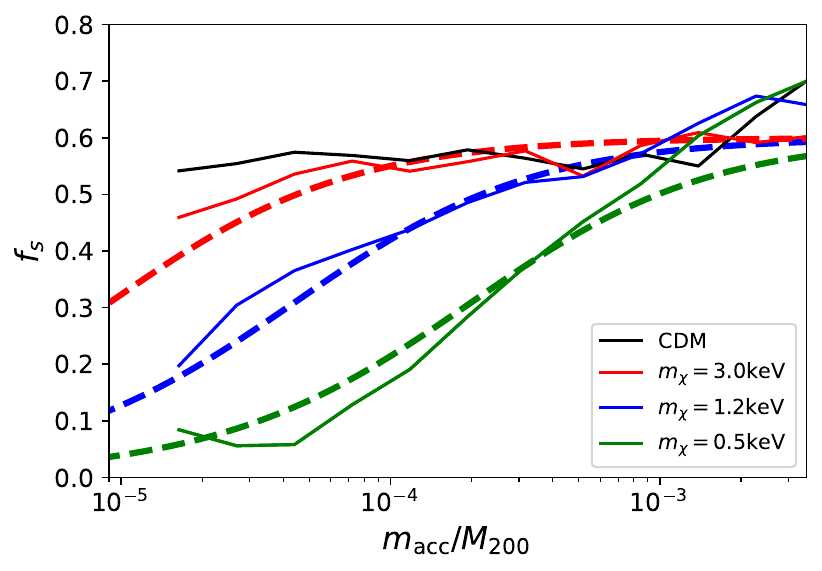}
    \caption{The survival rate of subhaloes as a function of the infall mass. The solid lines correspond to the simulation measurements as labelled. The dashed lines correspond to our analytical models.}
    \label{fig:fs}
\end{figure}

For the 1.2 keV WDM model, the survival fraction is very similar to that of the CDM case at the high infall mass bin. However, in contrast with the CDM case, a significant dependence on the infall mass can be observed. To account for this dependence, we model the survival fraction for WDM subhaloes as
%We obtain the survival fraction $f_s$ as a function of the infall mass of the subhalo in Fig.~\ref{fig:fs}. To model the dependence of $f_s$ on the WDM subhalo infall mass, we adopt the following form:
\begin{equation}
    f_s \left( m_{\mathrm{acc}} \right) = f_{s0} \left[ 1 + 3.81 (\frac{M_{200}}{m_0})^{-0.6}  \left ( \frac{m_{\mathrm{acc}}}{M_{200}} \frac{m_{\mathrm{acc}}}{M_{\mathrm{hm}}}  \right)^{-0.5} \right] ^{-0.85}.
\label{eq:Fs model}
\end{equation}
Here, $f_{s0}$ is the survival fraction in the CDM model. The model prediction is also shown in Fig.~\ref{fig:fs} with dash-dotted lines. It's worth noting that in Fig.~\ref{fig:lognormal }, we've binned the subhalo sample based on their infall masses larger than a given value, resulting in cumulative results $f_s(>m_{\rm{acc}})$, while Fig.~\ref{fig:fs} shows the differential result of $f_s$ as a function of the infall mass. %\jx{have you converted from the differential fs to $fs(>m_{\rm acc})$?} Compared with the CDM case, the WDM scenario leads to a higher fraction of subhalo disruption, especially at the low mass end. For a WDM particle mass of $0.5 \mathrm{keV}$, the survival fraction of subhaloes drops to nearly zero in the lowest infall mass bin. 

This infall mass dependence of $f_s$ is also required to explain the subhalo mass function. In the CDM case, both the unevolved and the final mass functions follow an approximate power law form, with an increasing number of subhaloes at lower masses. When subhaloes are stripped in a way independent of the infall mass, the final subhalo mass function can be approximately understood as a left-shifted version of the unevolved mass function. After the shift, the ratio between the two is also mass-independent as seen in Fig.~\ref{fig:dNdNacc}. % This ensures that the surviving subhaloes support the rising power-law mass function. Essentially, the final mass function can be considered as a left shift of the unevolved mass function. For a given final mass, the number of subhaloes is proportional to the unevolved mass function at higher masses. 
However, in the WDM scenario, the subhalo mass function is no longer scale-free but suppressed at the half-mode mass ($M_{\mathrm{hm}}$). If the final mass function is still a left-shifted version of the unevolved mass function, the ratio between the two cannot be mass-independent, in contradiction with Fig.~\ref{fig:dNdNacc}. This problem can be resolved when $f_s$ is mass-dependent, so that the abundance of lower mass subhaloes becomes further reduced. We will discuss this in more detail in section~\ref{sec:discussion}. % For instance, the number density of subhaloes with a final mass of $M_{\mathrm{hm}}$ is sourced from a larger infall mass bin in the unevolved mass function, where the value of the unevolved mass function is higher than that of the final mass function. Thus, the subhalo survival fraction at this mass should be smaller in the WDM model.

\section{Final model}\label{sec:results}

\begin{figure}
    \includegraphics[width=\columnwidth]{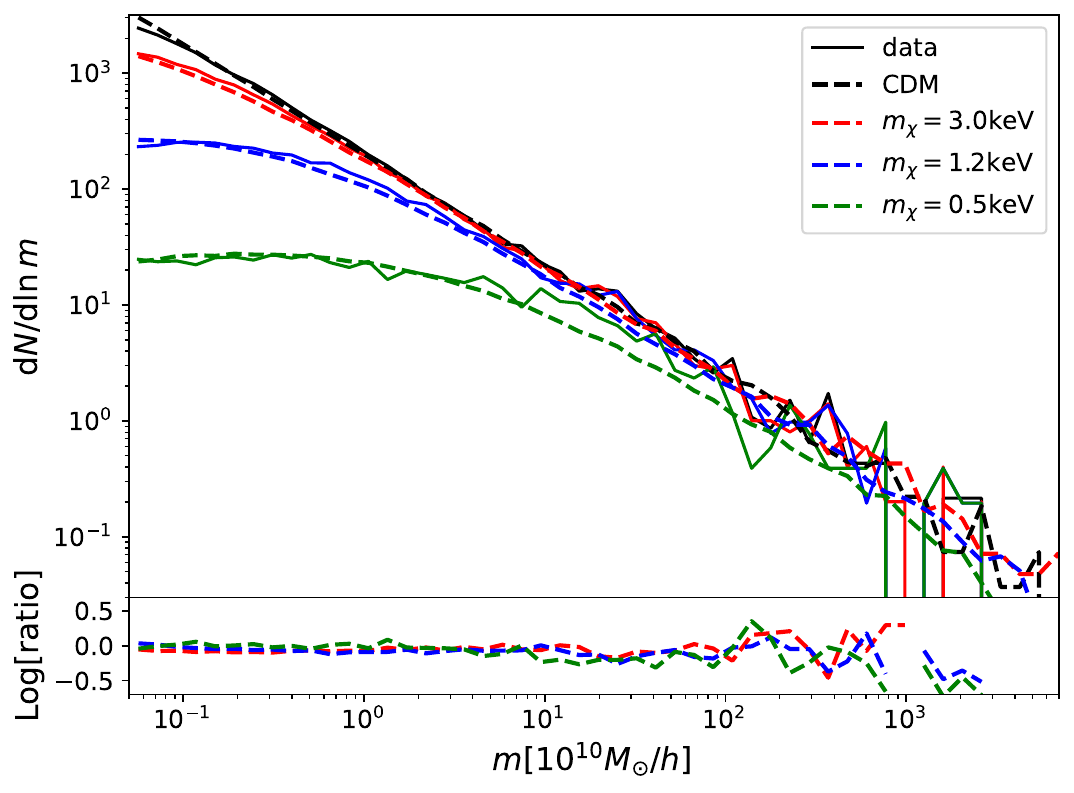}
    \caption{The subhalo mass functions from the simulation data (solid lines) and our model (dashed lines). The different colours correspond to different DM models as labelled. The bottom panel shows the ratio between our model predictions and the results from simulation data.} %\jx{remove ``model" from legend.}}
    \label{fig:SMF Model}
\end{figure}

\begin{figure}
    \includegraphics[width=\columnwidth]{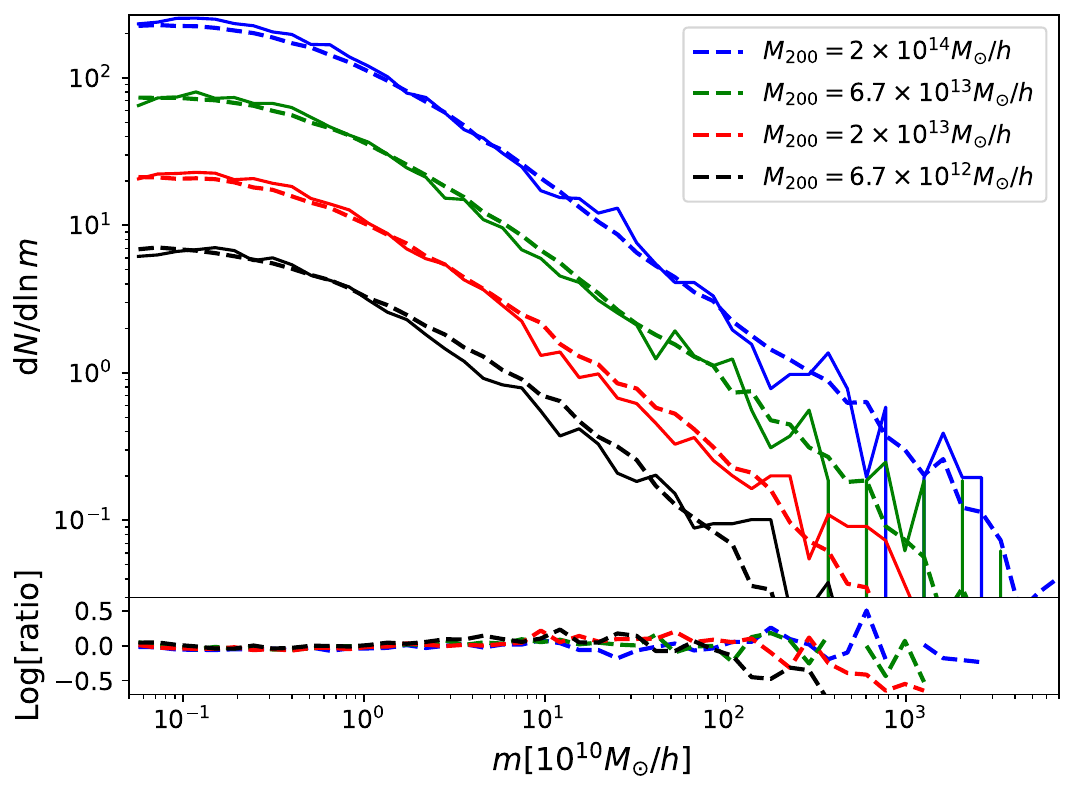}
    \caption{The 1.2keV WDM subhalo mass functions from simulation data (solid lines) and our model (dashed lines). The different colours correspond to different host halo masses as labelled. The bottom panel shows the ratio between our model predictions and the results from simulation data.}
    \label{fig:SMF on halo mass}
\end{figure}

In the previous section, we investigated the different model components in the WDM models, including the radial number density profile, the unevolved subhalo mass function, and the tidal stripping and disruption rate. Now we combine these components to build a complete model for the WDM subhalo distribution.

Following \citetalias{Han16}, the joint distribution of $m$, $m_{\mathrm{acc}}$ and $R$ is given by 
\begin{equation}
    \mathrm{d}N(m,m_{\mathrm{acc}},R) = \mathrm{d}N(m_{\mathrm{acc}})\tilde{\rho}(R)\mathrm{d}P(m|m_{\mathrm{acc},R}).\label{eq:fullpdf}
\end{equation} The three components are specified in section~\ref{sec:components}, with model parameters listed in Table \ref{tab:model_parameters_table}.

One can marginalize over the infall mass to get the final distribution of subhaloes as
\begin{equation}
    \begin{aligned}
        \frac{\mathrm{d} N(m, R)}{\mathrm{d} \ln m \mathrm{~d}^3 R} \sim & \tilde{\rho}(R)\int_{m_{\min }}^{m_{\max }} f_s \left( m_{\mathrm{acc}} \right) \left[\frac{m_{\mathrm{acc}}}{m_0}\right]^{-\alpha} \left(1+(a \frac {M_{\mathrm{hm}}} {m_{\mathrm{acc}}} )^{b}  \right)^{c}\\  & \times \exp \left[-\frac{1}{2}\left(\frac{\ln \mu-\ln \bar{\mu}(R)}{\sigma}\right)^2\right]  \mathrm{d}\ln m_{\mathrm{acc}} .
    \end{aligned}
\label{eq:joint distribution}
\end{equation}
Compared with the CDM case for which a simple approximate solution can be found, the integration of the above equation for WDM is more challenging. Alternatively, we can easily sample the full distribution specified by Equation~\eqref{eq:fullpdf}, and marginalize the Monte-Carlo sample over infall mass to obtain the final distribution corresponding to Equation~\eqref{eq:joint distribution}.

Fig.~\ref{fig:SMF Model} shows the subhalo mass function from the simulation data and our model predictions. The data is the average subhalo mass function in our selected cluster-size halo. From the CDM to different WDM models, the predictions from our models can match the simulation results very well.% \jx{do you have a quantitative measure of the difference?} %In addition, we also show the model without introducing the dependence of $f_s$ on $m_{\mathrm{acc}}$ in Fig.~\ref{fig:SMF Model nofs}. As discussed in Section.~\ref{sec:components}, if $f_s$ keeps constant in WDM cases, the evolved WDM subhalo mass function will have an offset from the correct position.

\begin{figure*}
\centering
 \includegraphics[scale=0.45]{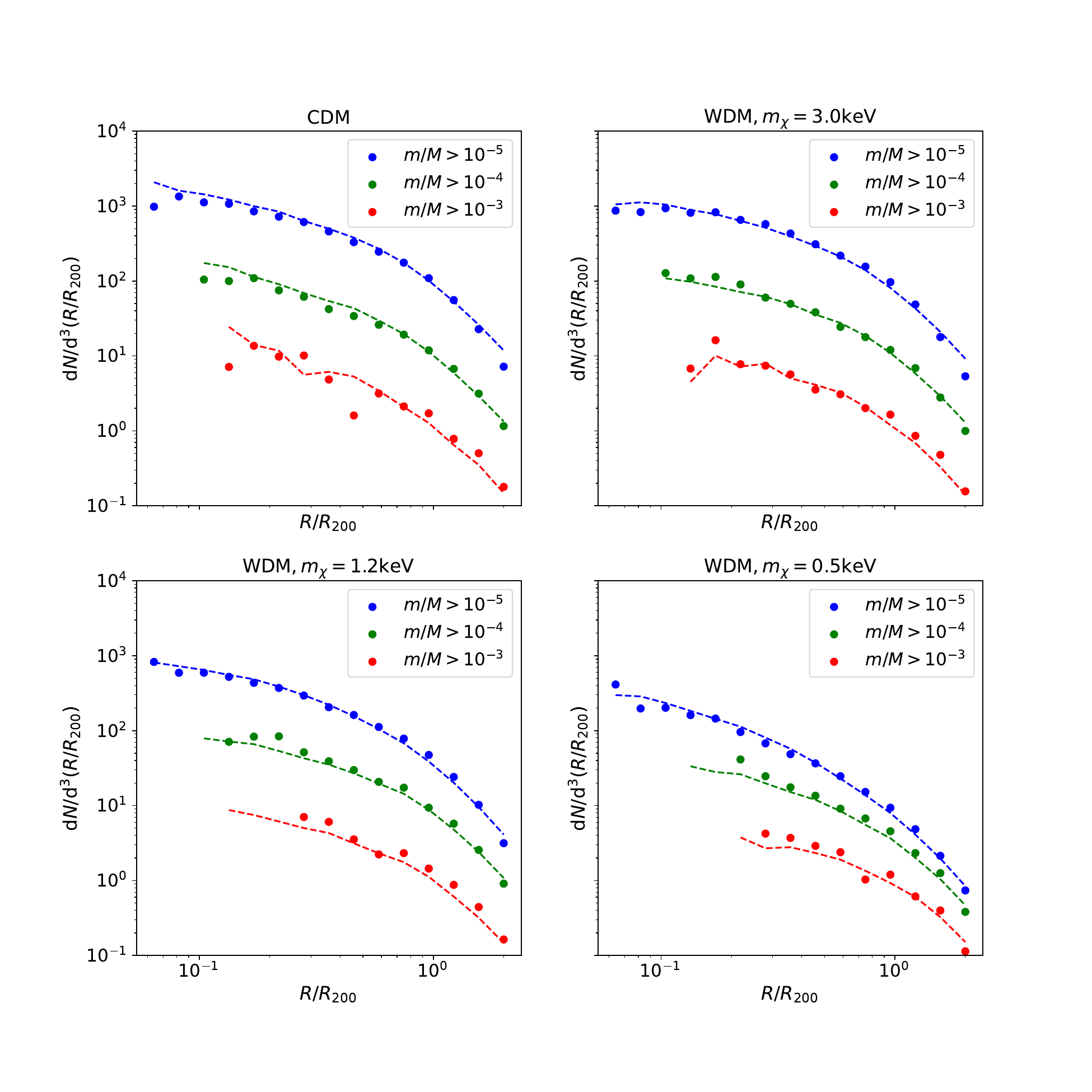}
 \caption{The radial number density profile of resolved subhaloes for a given final mass. The four panels correspond to the CDM subhaloes and the WDM subhaloes with $3.0\mathrm{keV}, 1.2\mathrm{keV}, 0.5\mathrm{keV}$. The dots are the results from the simulation data. The dashed lines show our model predictions. In each panel, from bottom to top, the three lines correspond to the subhaloes with $m / M_{200} > 10^{-3}, 10^{-4}, 10^{-5}$.}
 \label{fig:radial_distribution}
\end{figure*}

\begin{figure}
    \includegraphics[width=\columnwidth]{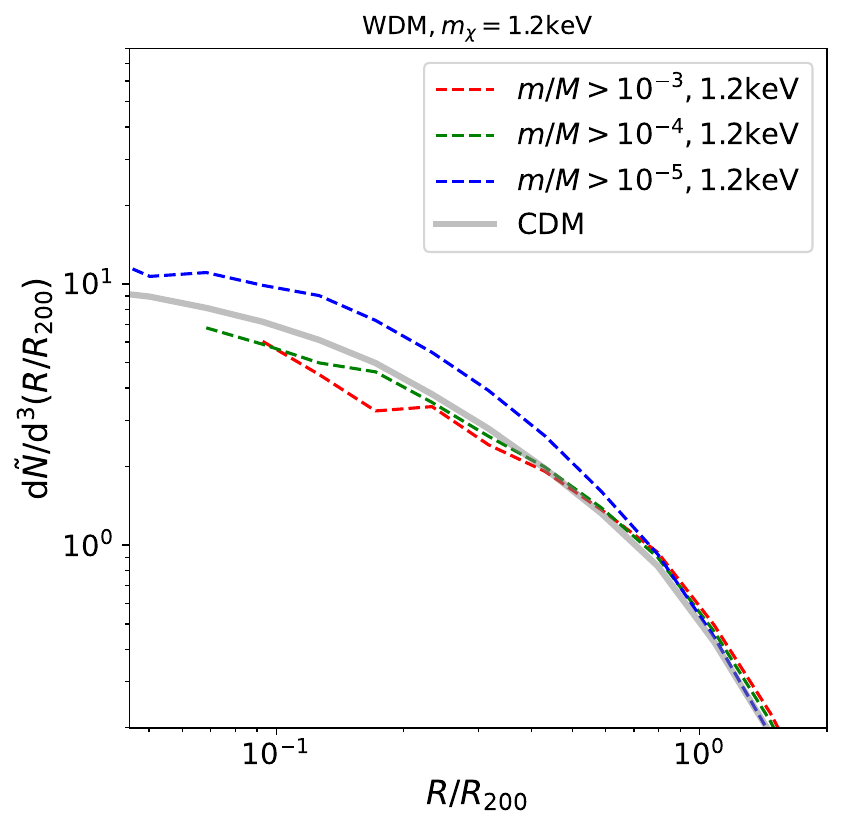}
    \caption{The scaled radial number density profile for 1.2keV WDM subhaloes with a host halo mass about $2\times 10^{14}h^{-1} M_{\odot}$. The dashed line shows our model predictions, corresponding to subhaloes with $\mu = m / M_{200} > 10^{-3}, 10^{-4}, 10^{-5}$ as labelled. The grey line corresponds to the scaled spatial distribution of CDM subhaloes, which is mass-independent. While the radial distribution of WDM subhaloes depends on the subhalo mass and the WDM particle mass.}
    \label{fig:radial_mass_dependence}
\end{figure}
 
So far we have only calibrated our model parameters on cluster-size haloes, which are not applicable to haloes of different masses. Given the limited resolution of our simulations, it is difficult to directly obtain the model parameters in lower-mass host haloes (such as MW-size haloes). Instead, we will assume the halo mass dependence of the WDM subhaloes is the same as that in the CDM case, and adopt the results from \citetalias{Han16} to model this dependence (see Table 1 in \citetalias{Han16}). This is equivalent to assuming that Equation~\eqref{eq: suppression function} and \eqref{eq:parameter_Mhm} are independent on host halo mass, while the halo mass dependence enters only through the CDM part. Note the $f_s$ parameter is independent on halo mass in the CDM case, compatible with Equation~\eqref{eq:Fs model}. With this final model, Fig.~\ref{fig:SMF on halo mass} compares the predicted subhalo mass functions in various host halo mass bins with our simulation results for the $1.2\rm{keV}$ WDM model, demonstrating the good agreement between our model and the simulation results.

Fig.~\ref{fig:radial_distribution} compares the spatial distribution of the resolved subhaloes in our simulation and model for a given final mass. In general, the spatial distribution given by our model can fit the results from the simulation. The amplitudes of different curves are determined by the subhalo mass function. For CDM subhaloes, the spatial distributions at different masses follow the same shape, as found in previous works~\citep{Springel08, Han16}. This distribution differs from the unevolved radial distribution due to mass stripping, as explained in Section~\ref{sec:framework}.

By contrast, the number density profile of WDM subhaloes is more complex and varies with the WDM particle mass, $m_{\chi}$. The distribution in the $m_{\chi}=3.0\rm{keV}$ WDM model is very close to that in the CDM case. With the decrease of $m_{\chi}$, the spatial distribution becomes more distinct. %The radial number density profile of the low-mass WDM subhaloes becomes steeper than that of the CDM subhaloes. 

\subsection{Key insights into the radial distribution of WDM subhaloes}
We show the scaled radial number density profile of $1.2\rm{keV}$ WDM subhaloes in different subhalo mass bins in Fig.~\ref{fig:radial_mass_dependence}. Compared with the CDM subhalo spatial distribution, the distribution of WDM subhaloes shows a clear dependence on the subhalo mass. The phenomenon has been discussed in \citet{2017MNRAS.464.4520B} following the theoretical framework of \citetalias{Han16}. According to the simplified formula of Equation.~\eqref{eq: CDM distribtuion}, the radial slope of the spatial distribution is determined by $\gamma=\alpha \beta$ and $\rho(R)$, as
\begin{equation}
    \frac{\mathrm{d} N(m, R)}{\mathrm{d} \ln m} \propto m^{-\alpha} R^\gamma \rho(R).
    \label{eq: simplified distribution}
\end{equation}
For the subhalo with a mass larger than $M_{\mathrm{hm}}$, the WDM subhalo mass function has the same slope $\alpha$ as that of the CDM. Below that scale, the shallower slope $\alpha$ leads to a shallower $\gamma$. Therefore, the radial number density profile for low-mass subhaloes should be less suppressed relative to $\rho(R)$, leading to a steeper radial distribution. 

\begin{figure*}
    \centering
        \includegraphics[scale=0.5]{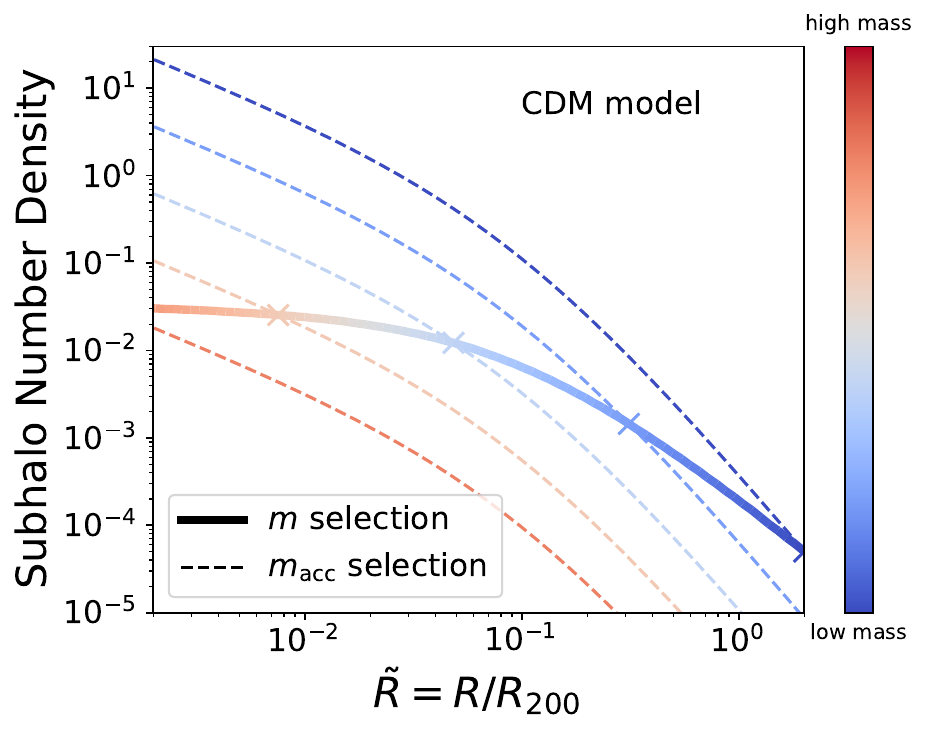}
        \includegraphics[scale=0.5]{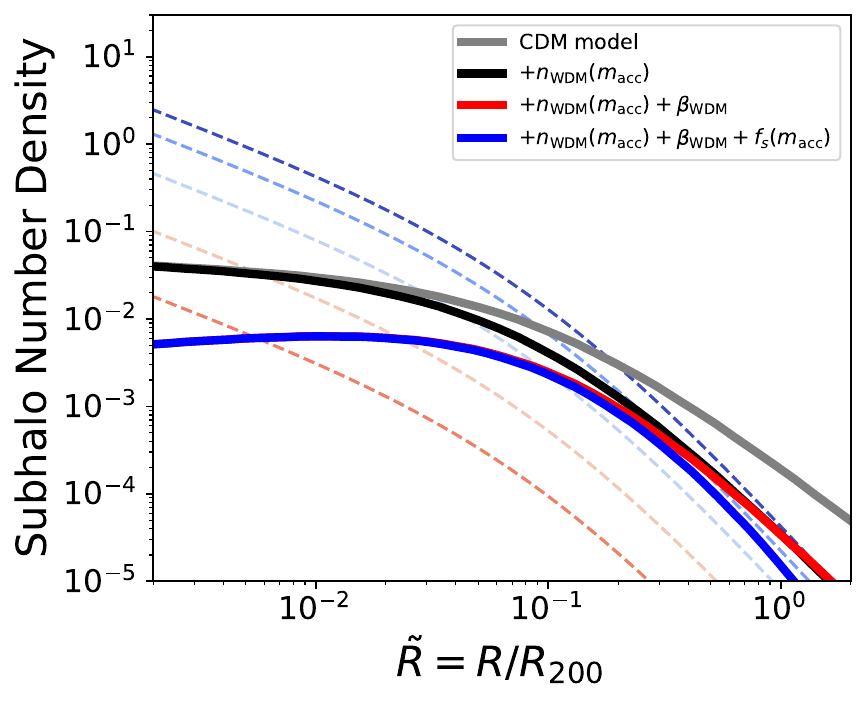}
    \caption{A toy model to demonstrate the subhalo number density profile within our model framework. The left panel illustrates the scenario in the CDM model. The thin coloured lines represent the unevolved radial number density distribution of subhaloes with varying infall masses. The colour bar indicates the corresponding changes in infall masses. The thick coloured line depicts the evolved radial number density profile of a subhalo sample with a given final mass. Cross points denote the contribution of the number density from subhalo samples of different infall masses to that of the given final mass subhalo sample at each radius. The right panel demonstrates how the final number density profile slope changes from the CDM case to the WDM case through the application of different WDM model components. The background dashed lines are analogous to those in the left panel but in the WDM scenario. The solid grey line represents the evolved subhalo number density profile given by the CDM model without any WDM model components. The solid black line showcases the result of incorporating the WDM suppressed unevolved mass function contribution into the CDM model. The subsequent red line incorporates a larger $\beta$  in the WDM model. Finally, the blue line represents the contribution of adding $f_s(m_{\rm{acc}})$.}
    \label{sub-radial-demo}
\end{figure*}

With our complete model, we can gain a more comprehensive understanding of the radial profile for WDM subhaloes. In our model, the difference between WDM and CDM subhaloes originates from three aspects, including the unevolved subhalo mass function, the tidal stripping law, and the survival rate. In Fig.~\ref{sub-radial-demo}, we explicitly show how these differences affect the resulting radial distribution of WDM subhaloes.%\jx{add the fig and finish this part}

To elaborate on these effects, we first review the origin for the radial distribution of CDM subhaloes in the left panel of Fig.~\ref{sub-radial-demo} following \citetalias{Han16}. Subhaloes of different infall masses all follow the same radial profile, but with different amplitudes as determined by the unevolved subhalo mass function. When selecting subhaloes of a given final mass, subhaloes of different infall masses are selected at different radii due to the different amounts of tidal stripping, with those at smaller radii having a larger infall mass. The number of subhaloes selected at the different radii then yields the solid curve in the figure, which is shallower in the inner part compared to the unevolved radial profile.

In the WDM case, three major modifications are introduced to the model of the final radial profile as shown in the right panel of Fig.~\ref{sub-radial-demo}. First, the unevolved radial profile curves are packed more tightly at the low mass end due to the suppression in the unevolved subhalo mass function. If we select subhaloes of a given final mass in the same way as in the left panel, the resulting radial profile (black solid curve) becomes steeper in the outer part compared to its CDM version (grey solid curve). Secondly, because the tidal stripping law is steeper in the WDM case ($\beta$ is larger), at a small radius the selected subhaloes have a larger infall mass than their CDM counterparts. This leads to a flatter inner profile as shown by the red solid curve, which also partly compensates for the steepening of the outer profile in the first point. Finally, because low infall-mass subhaloes have a smaller survival rate, the radial profile is further modified to be steeper in the outer part, as shown by the blue solid curve. The net effect of all three modifications leads to a radial profile that is steeper in the outer part and flatter in the inner part, with an overall lower amplitude compared to the CDM version. The transition between the flatter and steeper parts of the profile happens at a radius of 
\begin{equation}
    r_{\star}=\left( \frac{m}{10 M_{\rm{hm}} \mu_{*}} \right)^{1/\beta} R_{200},
\end{equation} where we have chosen a mass scale of $10 M_{\rm hm}$ to represent the transition mass in the unevolved mass function according to Fig.~\ref{fig:subhalo mass function suppression}. Note these three effects are simultaneously at play only in the modelling of low-mass WDM subhaloes. For high mass WDM subhaloes whose mass lie above $\sim 10M_{\rm hm}$, the first and third effect become irrelevant according to Equations~\eqref{eq: suppression function} and \eqref{eq:Fs model}, and we only expect an overall flattening of the radial profile due to the larger $\beta$.

\section{Discussions}\label{sec:discussion}

\subsection{Potential variations of the stripping parameters}
\begin{figure*}
    \includegraphics[scale=0.35]{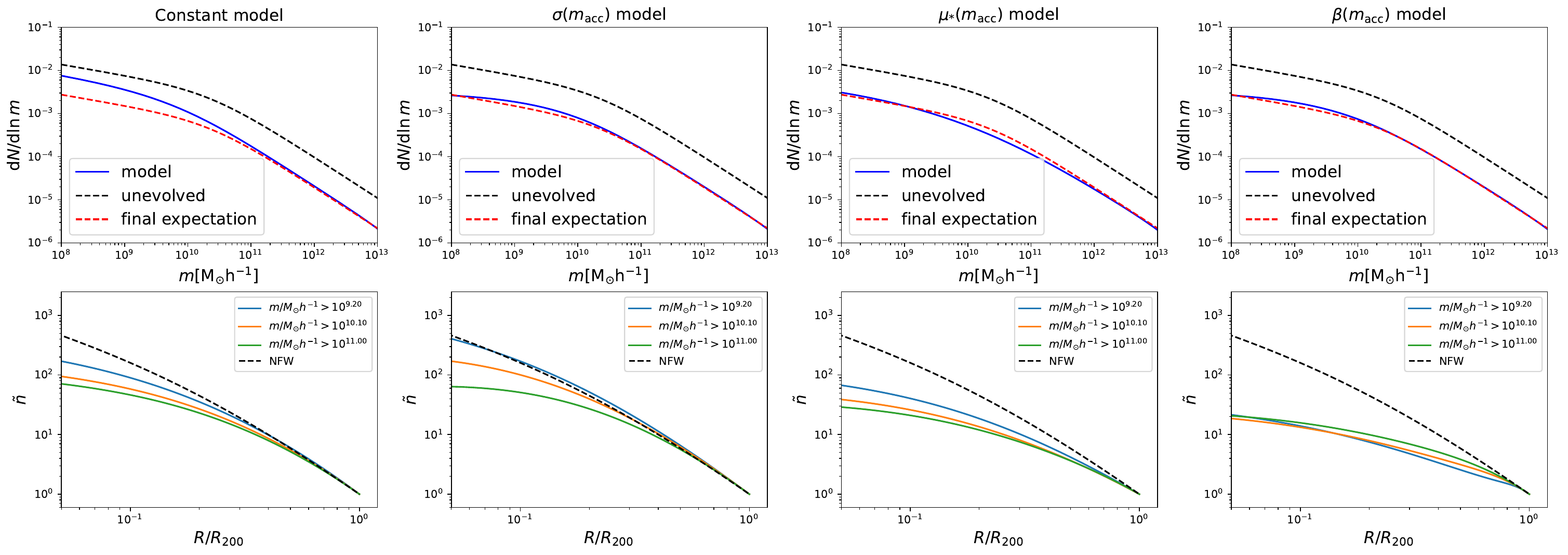}
    \caption{A toy model to illustrate how these parameters affect the final distribution of the subhalo. The left column shows the results of the constant model, in which all parameters are kept constant. The second column presents the results of the $\sigma(m_{\rm{acc}})$ model, where $\sigma$ has a dependence on the infall mass while other parameters remain constant. The third column presents the results of the $\mu_{*}(m_{\rm{acc}})$ model, where $\mu_{*}$ has a dependence on infall mass while other parameters remain constant. The right column displays the results of the $\beta(m_{\rm{acc}})$ model, where $\beta$ has a dependence on infall mass while other parameters are constant. The blue solid lines in the top row represent the subhalo final mass function of each model. The black dashed lines and red dashed lines in the top row correspond to the unevolved and expected final mass functions, respectively, as suggested in Fig.~\ref{fig:dNdNacc}. The bottom row exhibits each model's scaled subhalo radial number density profile, with different solid lines corresponding to the results of different mass bins. The dashed black lines in the bottom row represent the NFW profile.}
    \label{fig:numerical}
\end{figure*}

In Section~\ref{sec:results}, we have completed our model and compared it with the data. Due to the stronger tidal effects on the WDM subhaloes, the survival fraction will decrease with the subhalo infall mass. In \citetalias{Han16}'s model, Equation~\eqref{eq:strip_pdf} includes several parameters, $f_s,\beta,\sigma$, and $\mu_*$, describing the mass loss of a subhalo from the infall state to the final state. Except for $f_s$, do other parameters also depend on the infall mass? If there is a dependence on the infall mass, can the results of these models follow the data obtained from the simulations? In this section, we will address these problems through simple numerical integration based on our model framework. 

To avoid a complicated discussion, we divide the problems into four cases and deal with these parameters step by step. In each case, we set one parameter as a function of infall mass and other parameters as constants. As we have already studied the $f_s$ variation in our fiducial model, for the following we focus on generalizing the remaining three parameters. For simplicity, we set $f_s=1, M_{\rm{hm}}=10^{10}h^{-1} M_{\odot}, M_{200}=10^{14}h^{-1} M_{\odot}$ in these experiments. 

According to Fig.~\ref{fig:dNdNacc}, the ratio between the unevolved and final mass function is independent of the subhalo mass, which means the final mass function is expected to be simply a downward translation of the unevolved mass function. For each of our experiment, we try to make up a mass dependence of the corresponding parameter that can closely reproduce this expectation. The predicted mass and spatial distributions of subhaloes in all these cases are shown in Fig.~\ref{fig:numerical}.

We first test the case when all four parameters are independent of the infall mass. As shown in the top-left panel and discussed in section~\ref{sec:survival}, in this case, the predicted final mass function cannot be proportional to the unevolved mass function below the half-mode mass. 

Next, we test the mass dependence of $\sigma$, which is a parameter describing the dispersion of the bound fraction distribution. Here, we adopt a mass dependence of $\sigma(m_{\rm{acc}})=1+1.6(m_{\rm{acc}}/M_{\rm{hm}})^{-0.4}$, with which $\sigma$ rapidly increases when the infall mass becomes lower. With this model, the calculated subhalo final mass function is close to the expected result, as shown in the second column of Fig.~\ref{fig:numerical}. However, the model requires a large $\sigma$, which should be at least greater than 2.6 at $M_{\rm{hm}}$. In fact, Fig.~\ref{fig:lognormal } indicates that $\sigma$ is not sensitive to the infall mass, while it depends more on the WDM particle mass $m_{\chi}$. Even in the $0.5\rm{keV}$ WDM model, its $\sigma$ is only about 1.3, far from meeting the expectations of 2.6 expected in this model.

The second parameter to test is $\mu_{*}$, which is a parameter describing the dispersion of the bound fraction distribution. Here, we adopt a form $\mu_{*}(m_{\rm{acc}})=0.47(1 + 5(m_{\rm{acc}}/M_{\rm{hm}})^{-0.5} )^{-1}$, for which $\mu_{*}$ rapidly decreases when the infall mass becomes lower. After adopting this model, the calculated subhalo final mass function is close to the expected results, as shown in the third column of Fig.~\ref{fig:numerical}. However, the model requires a small $\mu_{*}$, which should be at least as low as 0.1 at $M_{\rm{hm}}$. In addition, we also do not find any evidence of its dependence on infall mass in Fig.\ref{fig:lognormal }.

The last parameter to test is $\beta$, which is the slope of the median mass loss over the halo-centric radius. Here, the adopted dependece is $\beta(m_{\rm{acc}})=1.3+2(m_{\rm{acc}}/M_{\rm{hm}})^{-0.5}$, for which $\beta$ rapidly increases when the infall mass becomes lower. As shown in the right column of Fig.~\ref{fig:numerical}, the final mass function can also be largely reproduced. However, the radial distribution of subhaloes calculated from this model shows a much shallower slope, which differs significantly from the simulation results.

To sum up, these experiments suggest that even though we cannot rule out the dependences of the other stripping parameters on the infall mass, it is unlikely for any of them to be the primary dependence in place of $f_s(m_{\rm acc})$.

\subsection{Reliability of subhalo disruption}

Recent studies indicate that the structure of the CDM subhalo in idealized N-body simulations is very robust, and displays strong resistance to complete disruption caused by tidal effects\citep[e.g.,][]{2018MNRAS.474.3043V,2020MNRAS.491.4591E,2020MNRAS.498.3902Y,2021MNRAS.505...18E}. Consequently, \citet{2018MNRAS.475.4066V} argue that subhalo disruption events observed in current state-of-the-art simulations might be artificial. The disruptions are mainly due to two numerical effects: one is the discreteness noise caused by the small number of particles in the system, and the other is the inadequate gravitational softening. 

However, it's important to note that their conclusion is only drawn from idealized numerical experiments. They adopted a series of experimental conditions, including an analytical static host potential, initial N-body subhalo systems generated through the Eddington inversion method, integration of motion along stable circular orbits, and minor mergers. In real cosmology simulations, the non-linear evolution undergone by a subhalo is more complex and it is difficult to directly meet the above experimental conditions, so whether their results are applicable to real cosmology simulations may need more detailed research. 

\citet{2021MNRAS.503.4075G} tested the effect of artificial disruption on the subhalo statistics using a semi-analytical model. They extracted a disruption model from the Bolshoi simulations and incorporated it into their semi-analytical model, \textsc{SatGen}. They found that the disruption events had an impact of $20\%$ on the final subhalo mass function at most. It is worth pointing out that the (sub)halo catalogue used in \citet{2021MNRAS.503.4075G} was found with \textsc{Rockstar}\citep{rockstar}, while our subhalo catalogue was constructed with the state-of-art tracking finder {\tt\string HBT+}. It has been known that the {\tt\string HBT+} (and its predecessor {\tt\string HBT}) is highly robust for tracking the evolution of subhaloes~\citep{2012MNRAS.427.2437H,2018MNRAS.474..604H}. Very recently \citet{2023arXiv230810926M} demonstrated that a particle-tracking method can be more robust compared to the popular combination of \textsc{Rockstar} and \textsc{ConsistentTrees}~\citep{ConsistentTree} in tracking subhaloes, enabling a more extended lifetime for the tracked subhalo. 

\citet{etheses15038} attempted a realistic cosmological test of the conclusions of \citet{2018MNRAS.475.4066V} by rerunning the Aquarius simulations. They extracted subhalo samples from Aq-A-2 halo and resimulated them using varying resolutions and gravitational softening. During this process, the potential of the host halo is represented analytically through the superposition of some basis functions~\citep{HEX}. They found that for the surviving subhaloes in their simulations, the softening length has no effect on whether the resimulated subhalo is disrupted or not, and different softening lengths eventually converge to the same mass loss fraction. The appearance of artificial disruption will only occur when the number of particles in the system is small enough. This provides further support to the reliability of our results on subhalo disruption statistics, which are based on the asymptotic distribution of the highly resolved subhalo population. In future works, we plan to carry out more comprehensive studies on the nature of subhalo disruption in cosmology simulations.

\section{Summary \& conclusions}\label{sec:summary}
In this work, we perform a series of DM-only simulations, including one cold dark matter model and three warm dark matter models with particle masses of $0.5\mathrm{keV}$, $1.2\mathrm{keV}$ and $3.0\mathrm{keV}$ respectively. The half-mode mass determined by the thermal relic WDM particle mass covers a wide range from $10^{8} h^{-1} M_{\odot}$ to $10^{11} h^{-1} M_{\odot}$. {\tt\string HBT+} is applied to identify subhaloes, followed by a cleaning procedure to remove spurious objects. Based on the cleaned WDM subhalo catalogues, we extend the unified subhalo distribution model of \citetalias{Han16} from CDM to WDM universes at $z=0$. 

The extension is done by adapting the three components of the model to appropriate forms for WDM subhaloes as follows.
\begin{enumerate}
\item Same as in the CDM case, the radial number density profile of subhaloes for a given infall mass follows the host halo density profile. We have verified that this is a fairly good assumption for the majority of subhaloes, although some deviations exist for massive subhaloes in the inner halo, reflecting the effect of dynamical friction.

\item Different from the CDM case, the unevolved subhalo mass function for WDM is no longer scale-free at the low mass end. Instead, there is a suppression in the mass function due to the small-scale cut-off in the initial power spectrum. 

\item The effect of tidal stripping is both stronger and more complex for WDM subhaloes. For surviving subhaloes, their distribution can still be modelled with a log-normal distribution in the bound mass fraction independent of the subhalo infall mass, although with different parameters from their CDM counterparts. On the other hand, the survival rate of subhaloes decreases with decreasing infall masses for WDM, in contrast to the mass-independent survival rate for CDM. 
\end{enumerate}
We use the half-mode mass of the initial power spectrum to universally parameterize the dependence of the model components on the dark matter particle property. This enables us to build a generalized model that can be applied to both CDM and WDM with different particle masses.

After calibrating the parameters of each model component separately, the final combined model successfully reproduces the final mass function and the spatial distribution of subhaloes in haloes of different masses and across all our simulations. The main conclusions from the model are the following.

\begin{itemize}
    \item Same as in the CDM case, the model predicts a final subhalo mass function that is proportional to the unevolved subhalo mass function for WDM. However, this proportionality is a consequence of both the scale-dependent unevolved mass function and the mass-dependent survival rate. By contrast, the proportionality in the CDM case is due to the scale-independent mass function along with a mass-independent tidal stripping law.

    \item In contrast to the spatial distribution that is independent of subhalo mass observed in CDM subhaloes, the radial distribution of WDM subhaloes does depend on the subhalo mass. Overall, less massive WDM subhaloes are more centrally concentrated than more massive ones. In addition, for low-mass WDM subhaloes, the radial number density profile is flatter in the innermost part but steeper outside when compared with its CDM counterpart. This complex radial profile is also a consequence of the scale-dependent mass function combined with the more complex tidal stripping law, as summarized in Fig.~\ref{sub-radial-demo}.
\end{itemize}

We have implemented our model into a python code, \textsc{SubGen2}, which can be used to populate haloes with subhaloes in the infall mass, final mass, and halo-centric radius, for both CDM and WDM cosmologies. Observationally, if we associate subhaloes of a given infall mass with galaxies of a given stellar mass following the subhalo abundance matching approach~\citep[e.g.,][]{2010MNRAS.402.1796W}, we can use this sampler to model the connections between galaxies and final their subhaloes, which can be tested by gravitational lensing studies~\citep[e.g.,][]{2014MNRAS.438.2864L,2016MNRAS.458.2573L, 2018MNRAS.475.4020W,2018MNRAS.478.1244S}. Future detections of not only luminous but also dark subhaloes through strong lensing~\citep[e.g.,][]{2009MNRAS.400.1583V,2016MNRAS.460..363L,2021MNRAS.506.5848E} or dynamical modelling of stellar streams~\citep[e.g.,][]{Carlberg12,Carlberg13,2019ApJ...880...38B} could also make use of our model to constrain the particle property of dark matter. Similar to what is done in \citetalias{Han16}, our model can also be adopted to predict indirect detection signals from the annihilation or decay of dark matter particles.

\section*{Acknowledgements}

We thank Ming Li for helpful discussions on using the \textsc{gadget4} code, and Mark Lovell for the generosity of offering access to the Aquarius-Warm simulations during the very early stage of this project. FH thanks Wenkang Jiang for the helpful discussions. This work is supported by NSFC (11973032, 11890691, 11621303), National Key Basic Research and Development Program of China (No. 2018YFA0404504), 111 project (No. B20019), and the science research grants from the China Manned Space Project (No.CMS-CSST-2021-A03). We thank the sponsorship from the Yangyang Development Fund. The computation of this work is partly done on the Gravity supercomputer at the Department of Astronomy, Shanghai Jiao Tong University.

%%%%%%%%%%%%%%%%%%%%%%%%%%%%%%%%%%%%%%%%%%%%%%%%%%
\section*{Data Availability}

The data underlying this article will be shared on a reasonable request to the corresponding author. The code for sampling subhaloes according to our generalized model is available at \url{https://github.com/fhtouma/subgen2}.

%%%%%%%%%%%%%%%%%%%% REFERENCES %%%%%%%%%%%%%%%%%%

\bibliographystyle{mnras}
\bibliography{example}

%%%%%%%%%%%%%%%%%%%%%%%%%%%%%%%%%%%%%%%%%%%%%%%%%%

%%%%%%%%%%%%%%%%% APPENDICES %%%%%%%%%%%%%%%%%%%%%

%%%%%%%%%%%%%%%%%%%%%%%%%%%%%%%%%%%%%%%%%%%%%%%%%%

% Don't change these lines
\bsp	% typesetting comment
\label{lastpage}
\end{document}